\documentclass[prb,twocolumn, superscriptaddress,showpacs, nofootinbib]{revtex4-2}
\usepackage{graphicx}
\usepackage{float}
\usepackage{dcolumn}
\usepackage{float}
\usepackage{bm}
\usepackage{mathrsfs}
\usepackage{ulem}
\usepackage[per-mode = fraction]{siunitx}
\usepackage{hyperref}
\hypersetup{
    colorlinks=true,
    linkcolor=blue,
    filecolor=magenta,      
    urlcolor=cyan,
    citecolor=magenta,
}
\usepackage{xcolor}
\usepackage{soul}
\usepackage{amsthm,amssymb}
\usepackage{mathtools}
\usepackage{comment}
\usepackage{physics}
\makeatother
\begin{document}

\title{Dielectric properties and plasmon modes of gapped momentum systems of different dimensionality}

\author{Yuriy Yerin}
\affiliation{CNR-SPIN, via del Fosso del Cavaliere, 100, 00133 Roma, Italy}
\author{A.A.~Varlamov}
\affiliation{CNR-SPIN, via del Fosso del Cavaliere, 100, 00133 Roma, Italy}
\author{Roberto Felici}
\affiliation{CNR-SPIN, via del Fosso del Cavaliere, 100, 00133 Roma, Italy}
\author{Aldo Di Carlo}
\affiliation{CHOSE (Centre for Hybrid and Organic Solar Energy), Department of Electronic Engineering, University of Rome Tor Vergata, Via del Politecnico 1, 00133 Rome, Italy}
\affiliation{Istituto di Struttura della Materia - CNR (ISM-CNR), EuroFEL Support Laboratory (EFSL), Via del Fosso del Cavaliere 100, 00133 Rome, Italy}

\date{\today}

\begin{abstract}
The concept of the energy gap is a fundamental characteristic of the band structure of a material  and it determines its physical properties. Formally the energy gap appears in the dispersion relation $E_k$, where the vector $k$ is determined on the whole momentum space. However, today the  {\it gapped momentum materials} are in the focus of research in which the so-called {\it momentum or $k$-gap} can emerge,  i.e. some lacunae of momentum space are excluded from the domain of the function $E_k$. One of such examples present the non-Hermitian systems. Within the random phase approximation we study the dielectric properties of the momentum gapped materials in one, two and three dimensions for both cases of zero and finite temperatures. We find the corresponding plasmon modes and determine the unusual behavior of the appropriate dispersion relations for each dimensionality. Based on these findings we evaluate the absorption coefficient of gapped momentum media and provide some numerical estimations of its value for the practical applications.
\end{abstract}
\maketitle

\section{Introduction}
The search for new materials with required properties for photovoltaics and similar applications is based on the detailed study of the band structure of the candidates. The analytical and numerical approaches based on microscopic formalism, such as the diagrammatic technique and DFT modeling, stand out in the first place for this purpose. The cornerstone of such methods is the study of the dispersion relation of electrons, which provides the dielectric properties, absorption, and valuable insights concerning possible topological features of electronic systems in the momentum space \cite{Bansil}. 

The materials with the parabolic dispersion relation have been extensively studied both analytically and numerically since 50s of the last century \cite{Ando, Giuliani}. Nowadays in purpose to reach the maximal efficiency of low-power-consumption electronics, quantum computations, photovoltaic technologies the new materials with the nontrivial topological band structure present a subject undergoing intense study.  These materials primarily include graphene, Weyl semimetals and Dirac gapped systems which dispersion relations are characterized by the root dependence on the square of the momentum modulus. The latter in the general form can be written as $E_{\bf k}=\sqrt{c^2{\bf k}^2+\Delta^2}$, where $\bf{k}$ is the momentum, $\Delta$ is the energy gap, while  $c$ is constant which can have various matrix structure, depending on the sort of material. As a result, such systems are characterized by the emergence of Dirac points or cones, Weyl nodes separated by the energy gaps in the momentum space. Obviously, the existence of such topologically exotic dispersion laws stimulates a theoretical study of their possible manifestation in the macroscopic characteristics of these materials in order to evaluate their potential for above mentioned practical applications.

\begin{figure}
\includegraphics[width=0.99\columnwidth]{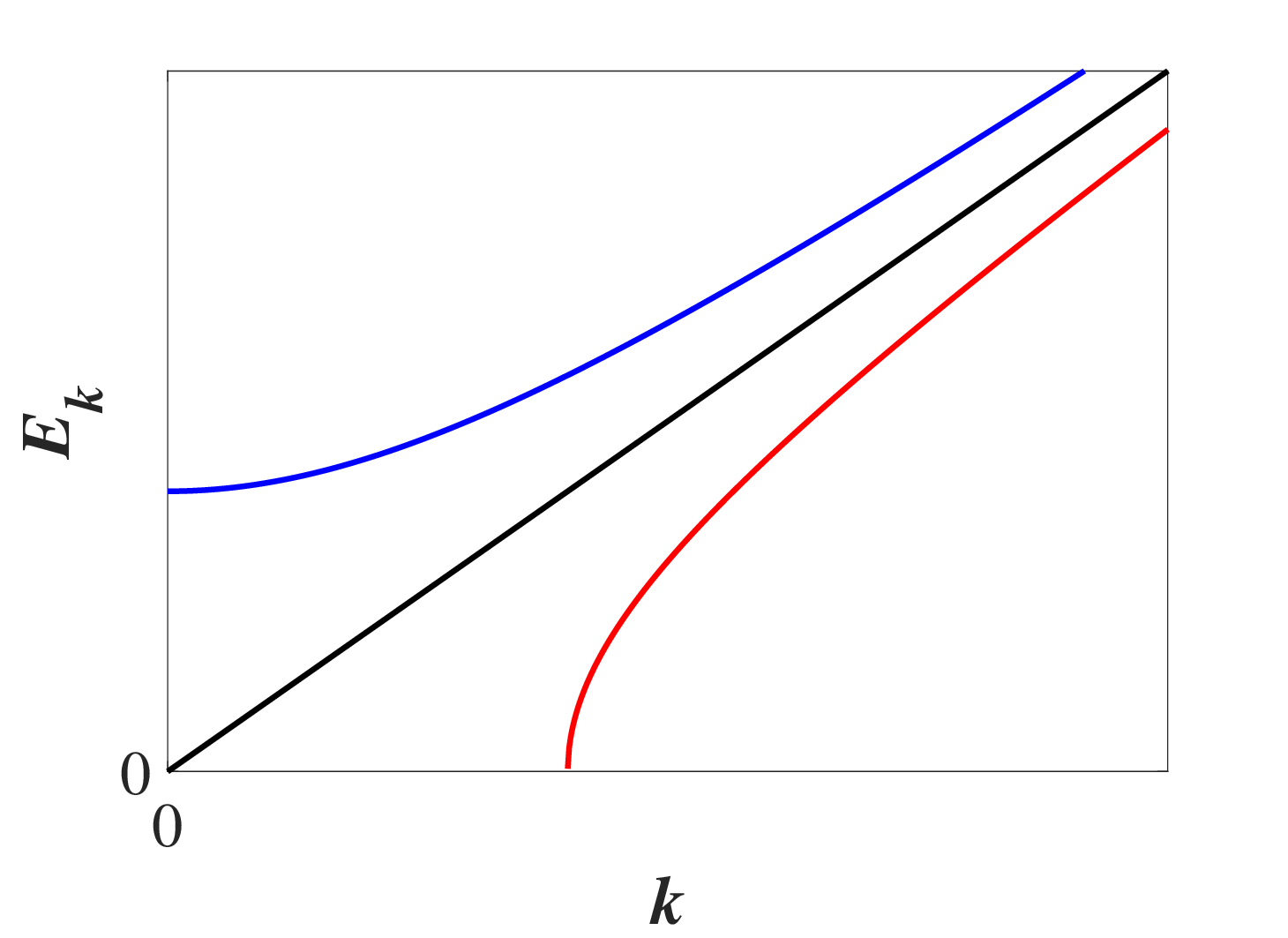}
\caption{Schematic illustration of different types of dispersion relations showing the dependencies of energy on momentum for Dirac gapped materials (blue line), graphene (black line) and the gapped momentum system (red line). The latter is symmetric to the top curve relative to the gapless black line.} 
\label{dr}
\end{figure}

Among the variety of dispersion relations already studied, it is worth noting another type, which is characterized by  emergence of the {\it momentum gap},  i.e., appearance of certain lacunae in momentum space which are excluded from the domain of the function $E_{\bf k}$ \cite{Baggioli1}. The example of a such dispersion relation with the momentum gap is shown in Figure \ref{dr}. Similarly, as a gap in the energy spectrum $\Delta$ indicates the absence of states between zero and the origin of the blue line on the axis $E_k$, so the momentum gap can be interpreted as the lack of momentum states between 0 and the origin of the black curve on the axis $k$. The presence of the momentum gap can be further interpreted differently in terms of the imaginary energy gap as $\Delta  \equiv i\Delta$. Such a concept has already been proposed in the context of the possible existence of particles known as tachyons, in which the energy gap plays the role of the mass with a complex value \cite{Feinberg}. Moreover, the possibility of the complex energy gap has recently attracted explosive attention to systems with non-Hermitian Hamiltonians and the emergence of a complex-energy spectrum, which can possess exceptional points (see e.g. review Ref. \onlinecite{Kawabata}). In particular, the emergence of a complex energy gap and consequently the possible realization of gapped momentum states are predicted in the non-Hermitian XY spin chain \cite{Liu}, the non-Hermitian BCS-BEC crossover of Dirac fermions \cite{Kanazawa}, the non-Hermtian graphene model \cite{Wu} and the non-Hermitian $\mathcal{PT}$-symmetric Fermi liquid \cite{Kruchkov}.

Hence, one can say that momentum gapped systems cannot be considered solely as the Gedankenexperiment. They appear also in a plenty of areas: ordinary liquids \cite{Trachenko1, Yang, Khusnutdinoff}, Dirac fluids \cite{Kiselev1}, charged two-dimensional liquids \cite{Kiselev2}, theory on the nonlinear elasticity \cite{Yerin}, strongly-coupled plasma \cite{Ohta, Murillo, Nosenko} (Ref. \onlinecite{Nosenko} is related to the experimental confirmation of the $k$-gap emergence in dusty plasma), anharmonic theory of superconductivity \cite{Setty}, sine-Gordon solitonic theory \cite{Kosevich} and holographic models \cite{Grozdanov, BT1, BT2, Gran, Arias}. 

For completeness of the picture, it is important to note that the dispersion relation describing gapped momentum systems may have even more complicated structure. In contrast to the case shown in Figure \ref{dr}, where the momentum gap appears at k=0 (red curve), there are systems in which the dispersion relation is characterized by one or more momentum gaps that emerge not only starting at $k=0$. For example, sound modes in fluids under specific conditions can show a discontinuous spectrum with several momentum gaps, also termed as propagation gaps in Refs. \onlinecite{Cohen, Zuilhof}. Despite this variety, in this paper we focus solely on the gapped momentum system with the dispersion relation illustrated schematically in Figure \ref{dr}.

However, to the best of our knowledge, the dielectric properties of the media with such a type of the dispersion relation still were not considered. Therefore, the goal of the present paper is to shed light on this topic and to study thoroughly a dielectric function of  a gapped momentum system in a one, two and three dimensional  cases. Since the possibility of practical application has been mentioned, knowing the dielectric function we will also quantify the absorption coefficient of gapped momentum media.

\section{Formalism and basic equations}

\begin{figure}
\includegraphics[width=0.49\columnwidth]{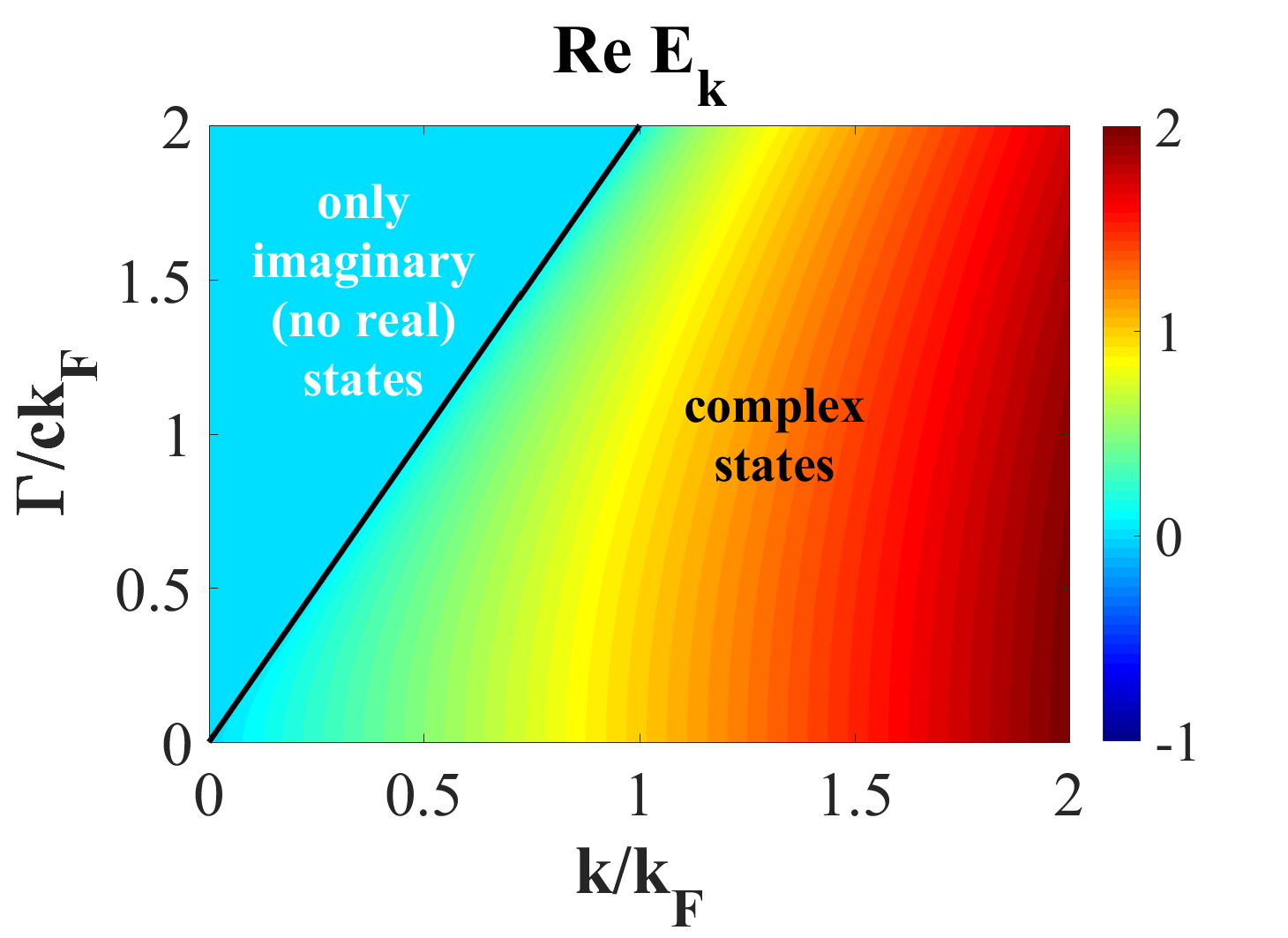}
\includegraphics[width=0.49\columnwidth]{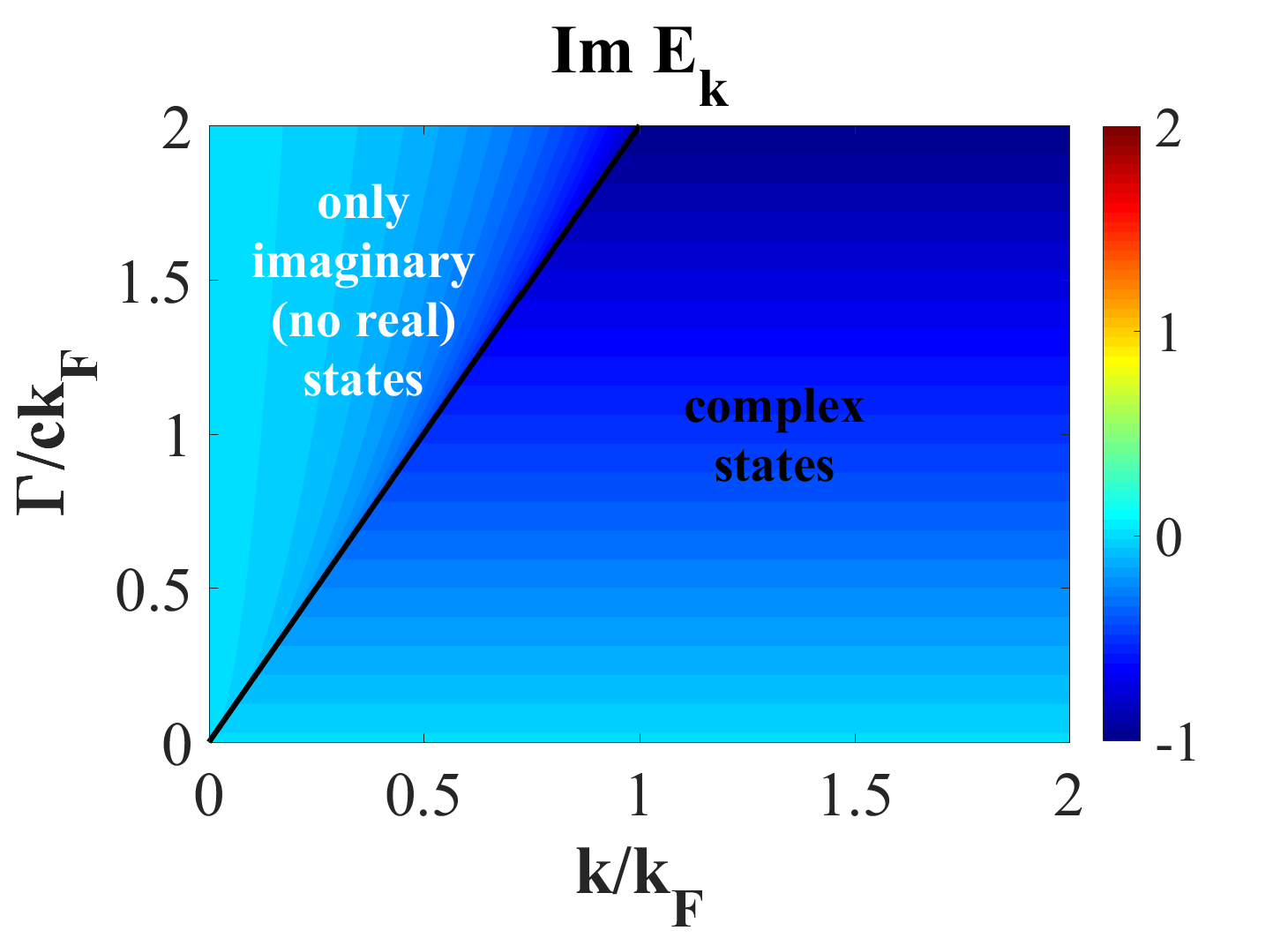}
\caption{Structure of the gapped momentum dispersion relation given by Eq. (\ref{dispersion_general})  in the complex plane $k$-$\Gamma$. The real (left) and the imaginary (right) part of the dispersion relation shows regions of where the real states are absence and the pure imaginary states occur and where they coexist all together. The black line serves as a boundary between these two regimes.} 
\label{dispersion_real_imag}
\end{figure}

There are several ways to justify the appearance of a gapped momentum dispersion relation (see e.g. reviews Refs. \onlinecite{Kovtun, Baggioli2}). We do not discuss the details of its derivation and provide the final expression written in the form of:
\begin{equation}
\label{dispersion_general}
{E_k} =  - \frac{{i\Gamma }}{2} + \sqrt {{c^2}{k^2} - \frac{{{\Gamma ^2}}}{4}}.
\end{equation}
Here $k$ is the momentum of a electron, $\Gamma$ and $c$ are the certain positive constants. We are agnostic about their microscopic origin, however the coefficient $\Gamma$ can be interpreted as the damping term related to the inelastic interactions, or as a complex energy gap $\Gamma=i\Delta$ in non-Hermitian systems (see Introduction). The coefficient $c$ in the simplest case of gapped Dirac material is nothing else as the Fermi velocity $v_F$. Figure \ref{dispersion_real_imag} focuses on the real and the imaginary part of the dispersion relation Eq. (\ref{dispersion_general}) as a function of $k$ and $\Gamma$. These contours plot clearly indicate the absence of states, when $ck<\Gamma/2$ and their formal emergence in the complex momentum space.

\subsection{Dielectric function.} 
We start the calculation of the dielectric function $\varepsilon \left( {q,\omega } \right)$  for the system with the gapped momentum  dispersion in the framework of the random phase approximation (RPA) \cite{Mahan}:
\begin{equation}
\label{epsilon_general}
\varepsilon \left( {q,\omega } \right) = 1 - {V_q}\Pi \left( {q,\omega } \right),
\end{equation}
where $\Pi \left( {q,\omega } \right)$ is the retarded polarization function of the non-interacting electron system described by the spectrum (\ref{dispersion_general}), which in turn can be written as
\begin{equation}
\label{Polarization_general}
\Pi \left( {{\bf{q}},\omega } \right) = \sum\limits_{\bf{k}} {\frac{{f\left( {\bf{k}} \right) - f\left( {{\bf{k}} + {\bf{q}}} \right)}}{{\omega  + {E_{\bf{k}}} - {E_{{\bf{k}} + {\bf{q}}}} + i\eta }}}.
\end{equation}
Here we denote  $\bf{q}$ as the wave vector, $\omega$ as the real frequency, and $\eta \to 0$. The nominator of Eq. (\ref{Polarization_general}) represents the difference of Fermi-Dirac functions for different momenta $\bf{k}$ and $\bf{k=k+q}$. The summation (integration) over $\bf{k}$ is performed with the upper limit equal to the Fermi momentum ${k_F} = \frac{1}{c}\sqrt {{\mu ^2} + \frac{{{\Gamma ^2}}}{4}}$, where $\mu$ is the chemical potential, which is assumed to be positive.  $V_q$ is the Fourier transform of the corresponding bare Coulomb interaction for the given dimensionality $d$ of a system given by the expressions
\begin{equation}
\label{V_q}
{V_q} = \left\{ \begin{array}{l}
2\pi {e^2}{K_0}\left( {aq} \right), \ {\rm{ for \ 1D}},\\
\frac{{2\pi {e^2}}}{q}, \hspace{1.36cm} {\rm{ for \ 2D}},\\
\frac{{4\pi {e^2}}}{{{q^2}}}, \hspace{1.36cm} {\rm{ for \ 3D,}}
\end{array} \right.
\end{equation}
where $K_0(x)$ is the zeroth order modified Bessel function of the second kind and the parameter $a$ corresponds to the length scale, which characterizes the lateral confinement size of the 1D geometry. For the sake of completeness, we will study all three above mentioned cases to elucidate the effect of dimensionality on the properties of the dielectric function.

In the polarization function Eq. (\ref{Polarization_general}) we ignore the electron-hole interactions and transitions between the conduction (electron) and valence (hole) bands like in Dirac gapped materials \cite{Gorbar, Pyatkovskiy, Thakur} and Weyl semimetals \cite{Lv} usually described by introducing of the corresponding overlap function.

It is worth noting that the theoretical justification of gapped momentum states can be developed also in the framework of non-Hermitian two-field (with two scalar fields) theory with broken $\mathcal{PT}$ symmetry \cite{Tkachenko2,  Tkachenko3}. Such a field theory formulation includes dissipation process, and hence requires considering quantum dynamics of the system out-of-equilibrium.  This means that in this case it is necessary to re-derive RPA taking into account non-equilibrium effects in the framework of the Kadanoff-Baym-Keldysh formalism. In this paper we are limited to the simplest version of RPA, which is a standard approximation commonly used to obtain the exchange and correlation energies of an equilibrium free electron gas. Therefore, the inclusion of non-equilibrium effects is a separate problem, which is beyond the scope of our paper and will be subject of future studies. 

 \subsection{Plasmon dispersion.} 
 The plasmon dispersion relations within the RPA are determined by the zeros of the dielectric function Eq. (\ref{epsilon_general})
\begin{equation}
\label{plasmon_general}
1 - {V_q}\Pi \left({ {\bf{q}},\omega }\right) =0.
\end{equation}

Obviously, in general case there is no analytical solutions of Eq. (\ref{plasmon_general}). However, at $T=0$ and in the long-wave limit when $q \to 0$ one can reduce the expression for the polarization function Eq. (\ref{Polarization_general}) to
\begin{equation}
\label{Polarization_simple}
\Pi \left({ {\bf{q}},\omega }\right)  =\sum\limits_k {\frac{{{{\bf{v}}_k} \cdot {\bf{q}}}}{{\omega  - {{\bf{v}}_k} \cdot {\bf{q}} + i\eta }}} \delta \left( {{E_k} - \mu } \right),
\end{equation}
where ${{{\bf{v}}_k}}={\nabla_{\bf k} {E_k}}=\frac{k}{{\sqrt {{k^2} - \frac{{{\Gamma ^2}}}{4}} }}$ is the group velocity and $\delta(x)$ is the Dirac delta-function. Noteworthy, at $k= \Gamma/2$ the border between purely imaginary and complex (mixed real and imaginary) states is characterized by an infinite velocity group. However, it was shown that such a divergence indicates the occurrence of non-propagating excitations and hence no superluminal effects are present \cite{Baggioli1}. In any case this does not impact the analytical calculations because, as can be seen from Eq. (\ref{Polarization_simple}), this point is a removable singularity for the expression of the polarization function.

Then, in order to find the long wavelength plasmon dispersion, one can apply series expansion of the denominator in Eq. (\ref{Polarization_simple}) over $\bf{q}$ and obtain \cite{Grecu}
\begin{equation}
\label{plasmon_approx_eq}
{V_q}\sum\limits_k {\frac{{\nabla_{\bf k}{E_k} \cdot {\bf{q}}}}{\omega + i\eta }\left( {1 + \frac{{\nabla_{\bf k}{E_k} \cdot {\bf{q}}}}{\omega + i\eta } + ...} \right)} \delta \left( {{E_k} - \mu } \right) = 1.
\end{equation}
It is this equation that will be used to calculate analytically the plasma frequency $\omega_{pl}$ in the following sections.

\section{The dielectric function and plasmon modes}

In general, the dielectric function $\epsilon(q,\omega)$ is the basis for understanding and predicting a plethora of static and dynamic many-body effects in various electronic systems. In particular, the static dielectric function $\epsilon(q,\omega=0)$ can be used for studying of Friedel oscillations and the Kohn anomaly (that normally occurs when $q=2k_F$) to understand new features caused by the unusual dispersion relation. Since here we are concentrated on the possible photovoltaic applications of a gapped momentum material, we will focus our attention mainly on the frequency dependence of the dielectric function and plasmonic modes within the long-wavelength approximation. Moreover, unlike many previous works, we extend our investigation of the behavior of the dielectric function for nonzero temperatures.

\subsection{1D gapped momentum material}

\subsubsection{The dielectric function}
Although the one-dimensional case allows the straightforward analytical calculation of the dielectric function, the integration of Eq. (\ref{Polarization_general}) yields a very cumbersome expression inconvenient for further analysis even for zero temperature. In particular, at $T=0$, when one can replace the Fermi-Dirac distribution by the Heaviside function, the final answer for the polarization function is expressed in terms of the elliptic integrals of the first, second and third kind, and do not lead to any useful insight (see Appendix \ref{sec:A}, where we provide the exact analytical expression for $\Pi \left( {{\bf{q}},\omega } \right)$). Nevertheless, this analytical expression will serve as a verification of our subsequent numerical calculations.

Before proceeding to numerical calculations of $\varepsilon \left( {q,\omega } \right)$ it is important to note that the RPA approach works significantly better in higher dimensions as opposed to 1D case, where the Luttinger liquid theory \cite{Giamarchi} is necessary to apply for the description of the interacting itinerant electrons. However, RPA method in this case allows to find with a good accuracy the plasmon mode. This fact is evidenced by the satisfactory coincidence of the RPA results for the plasmon modes in the 1D electron system with the parabolic dispersion and the corresponding experimental observations in the long-wavelength limit \cite{Goni}.

\begin{figure}
\includegraphics[width=0.49\columnwidth]{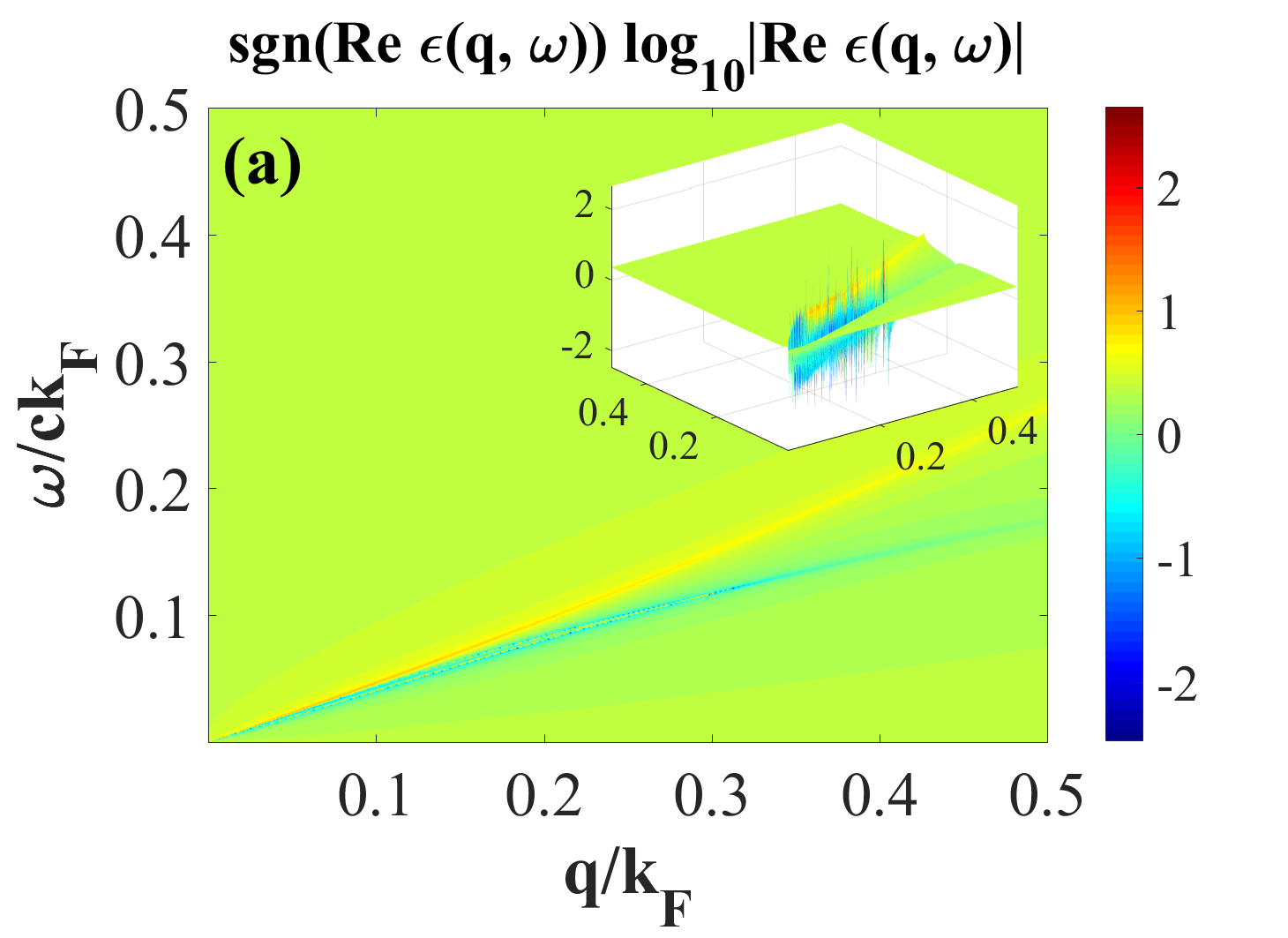}
\includegraphics[width=0.49\columnwidth]{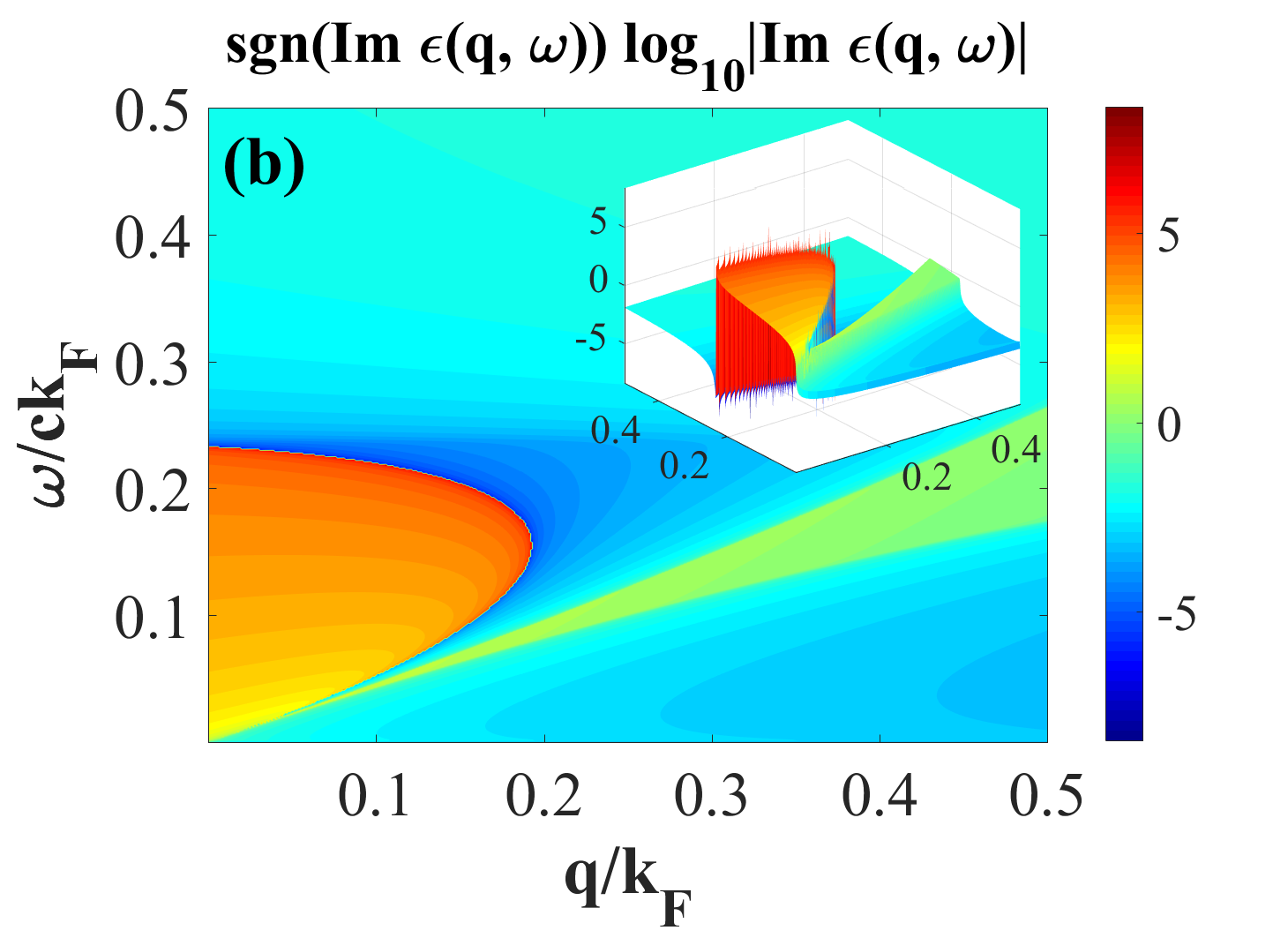}
\includegraphics[width=0.49\columnwidth]{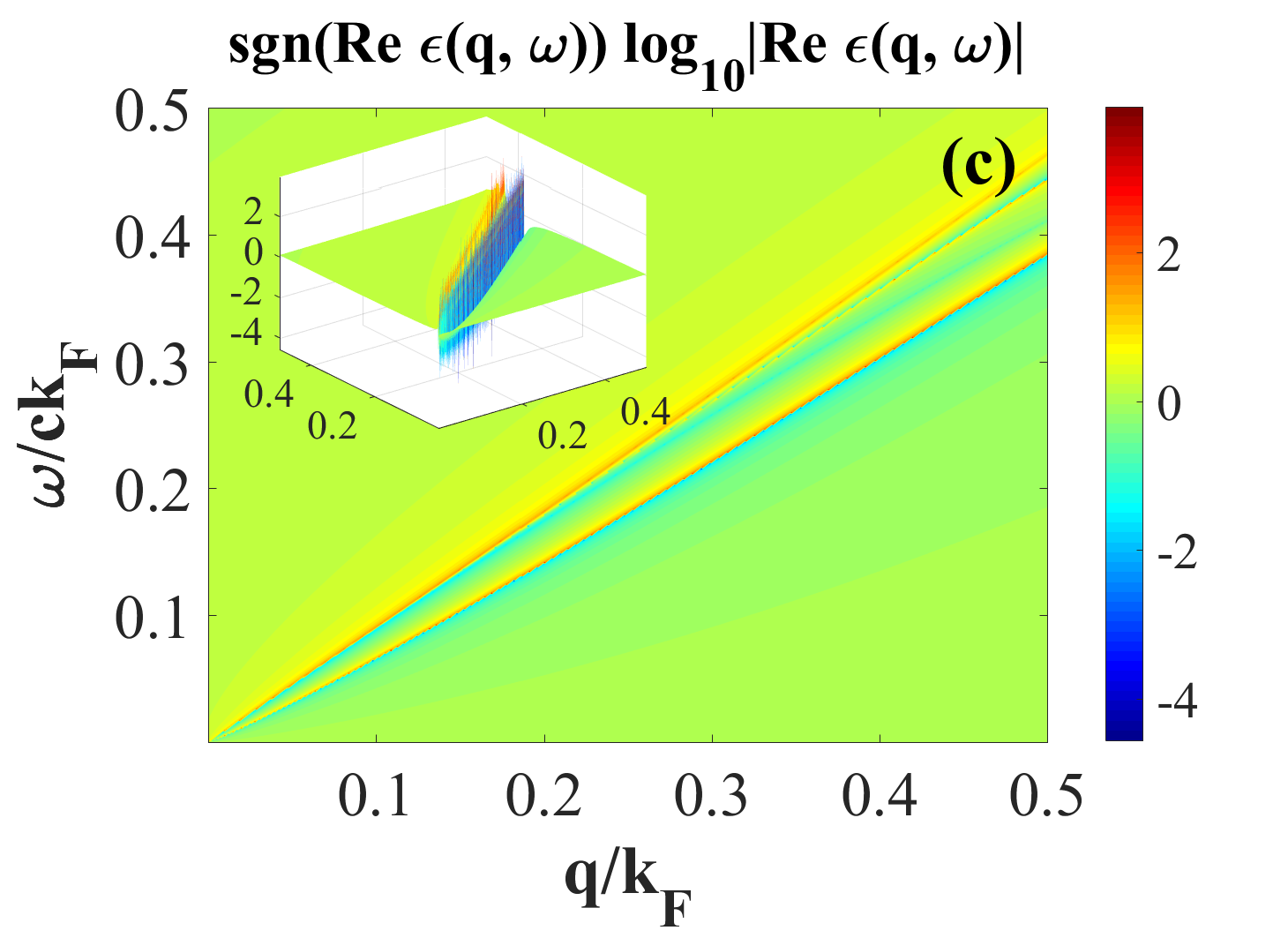}
\includegraphics[width=0.49\columnwidth]{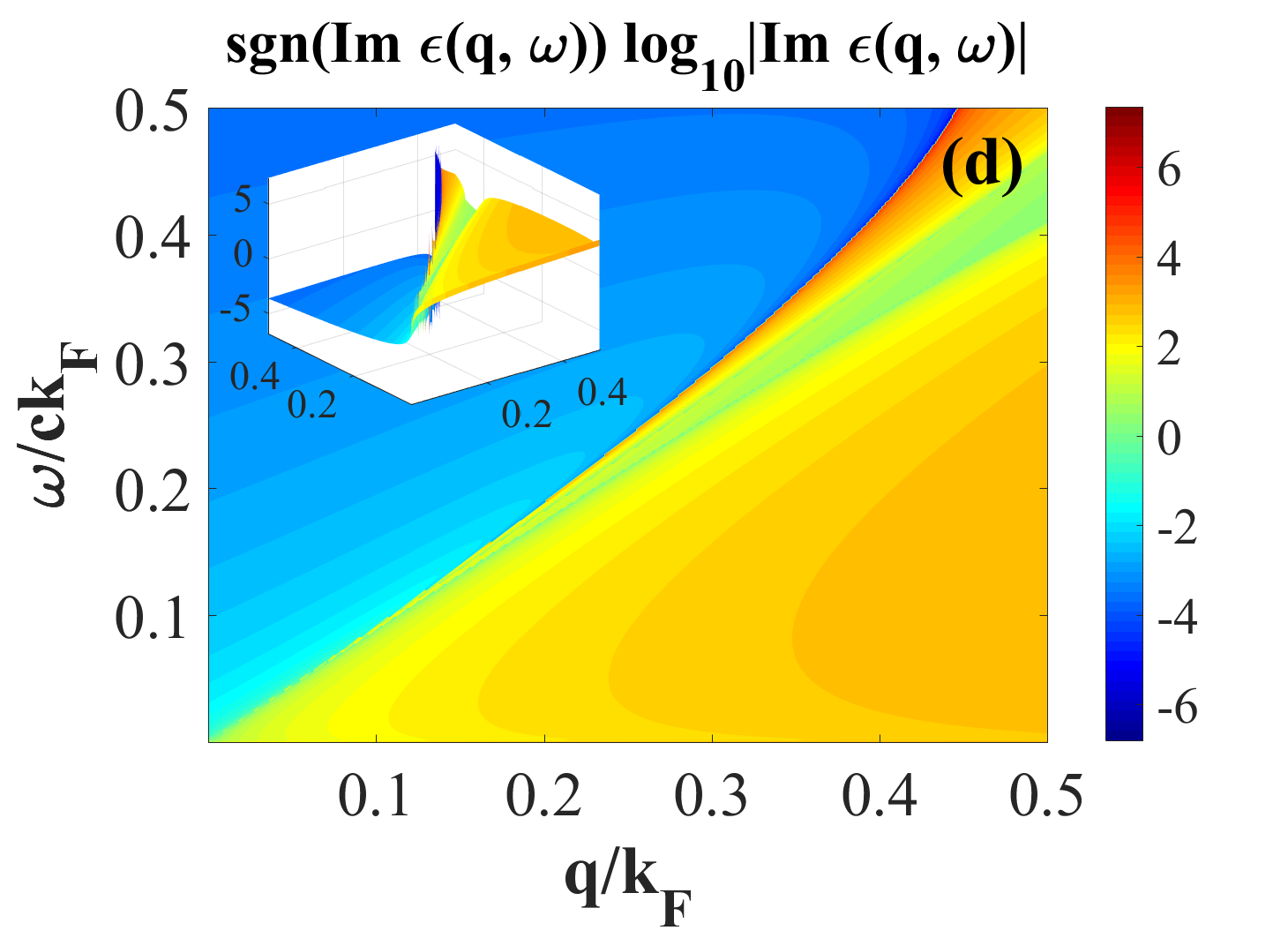}
\caption{The real (a, c) and imaginary (b, d) part of the dielectric function $\epsilon(q,\omega)$ in $q$-$\omega$ plane for a 1D gapped momentum material with the parameter $\Gamma/ck_F=4$ (a,b) and $\Gamma/ck_F=1$ (c,d) and for the temperature $T=0$. For the sake of clarity here and hereafter, we plot the dielectric function as ${\mathop{\rm sgn}} \left( {{\mathop{\rm Re}\nolimits} \ \varepsilon \left( {q,\omega } \right)} \right){\log _{10}}\left| {{\mathop{\rm Re}\nolimits} \ \varepsilon \left( {q,\omega } \right)} \right|$ and ${\mathop{\rm sgn}} \left( {{\mathop{\rm Im}\nolimits} \ \varepsilon \left( {q,\omega } \right)} \right){\log _{10}}\left| {{\mathop{\rm Im}\nolimits} \ \varepsilon \left( {q,\omega } \right)} \right|$. Insets show three-dimensional surface plot of $\epsilon(q,\omega)$ also in the logarithmic scale to illustrate additionally the behavior of the dielectric function, which is not clear visible in contour plots in the region of small $q$ and $\omega$.} 
\label{epsilon1D_T=0-001}
\end{figure}

\begin{figure}
\includegraphics[width=0.49\columnwidth]{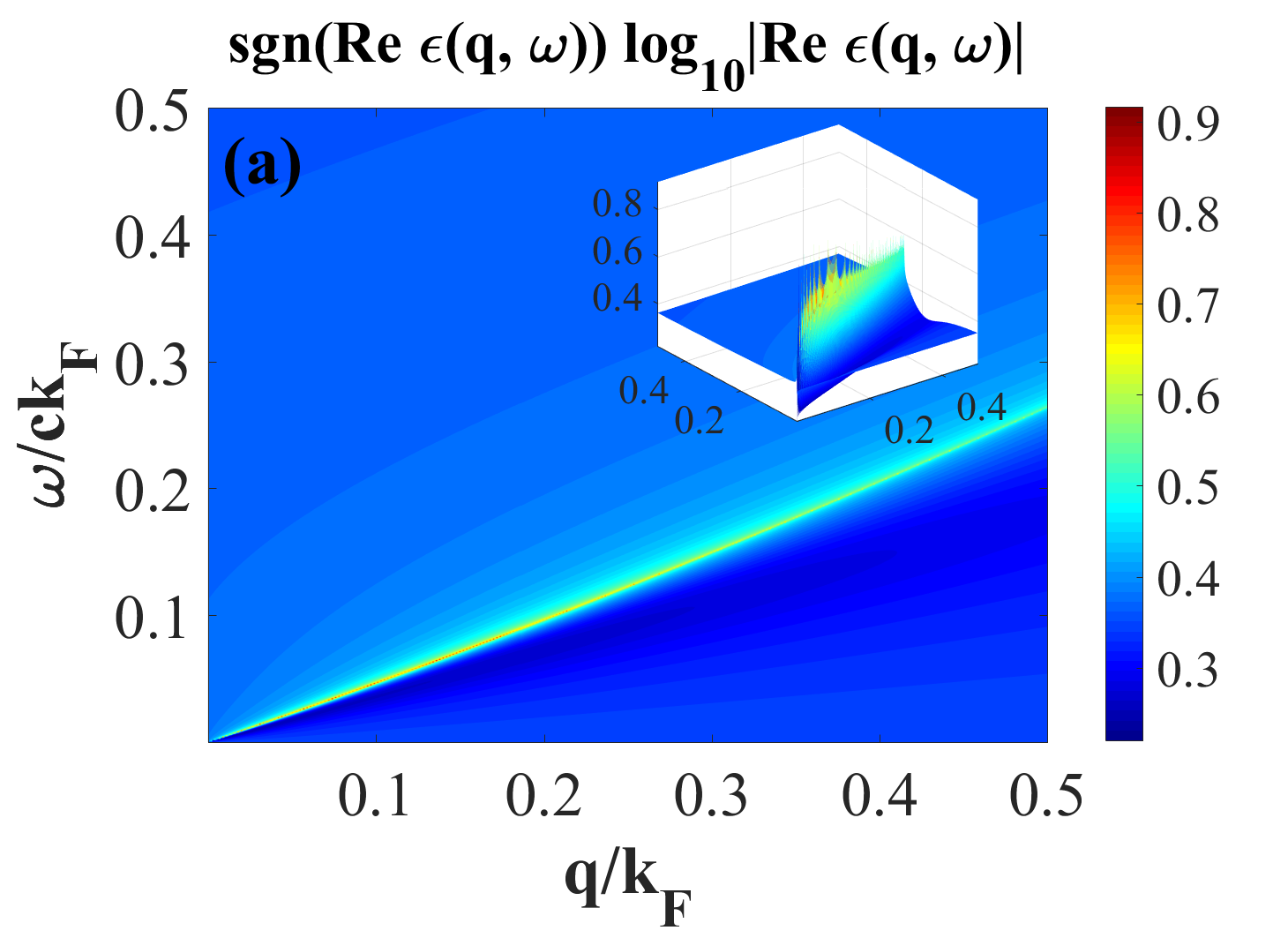}
\includegraphics[width=0.49\columnwidth]{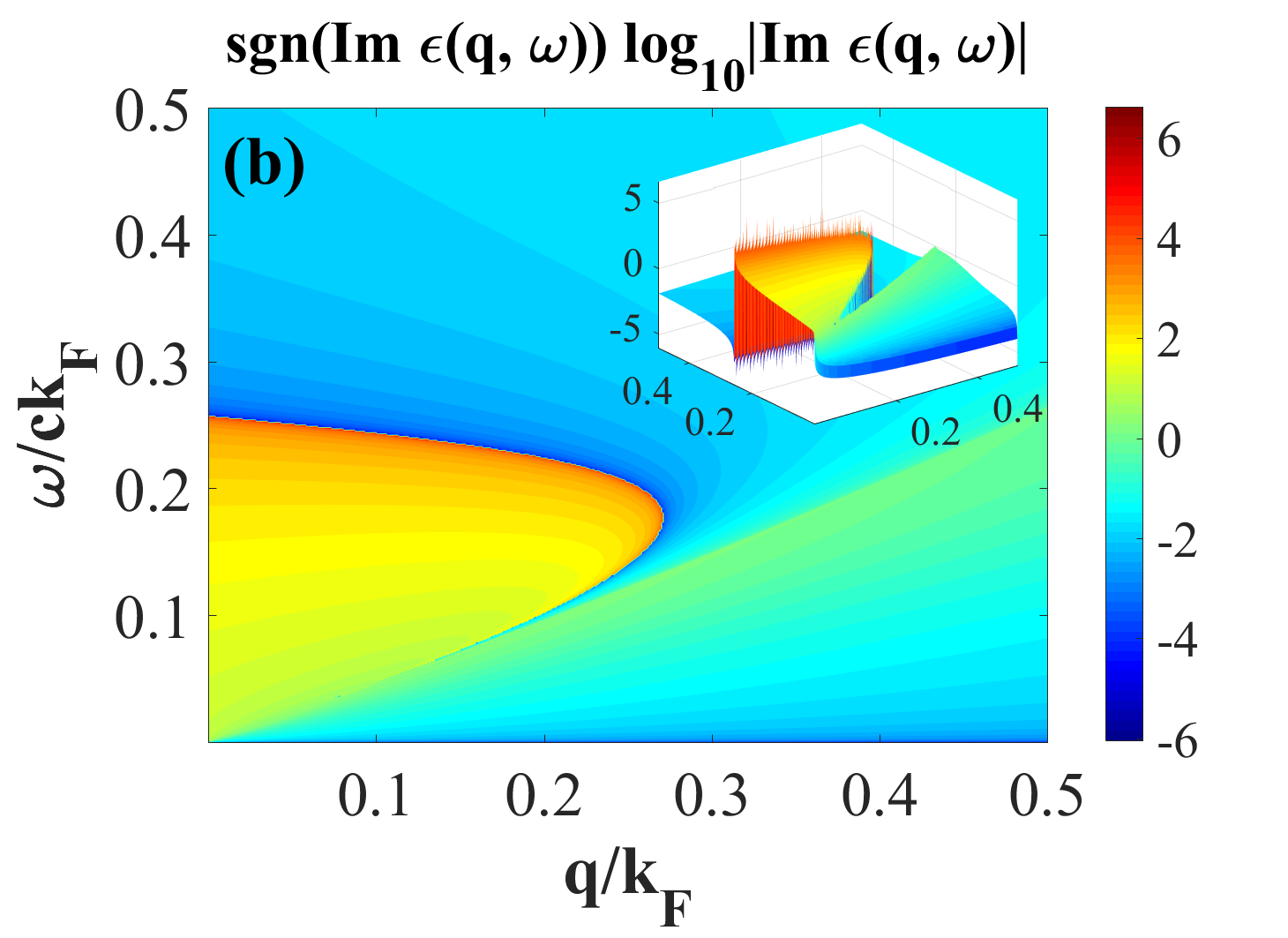}
\includegraphics[width=0.49\columnwidth]{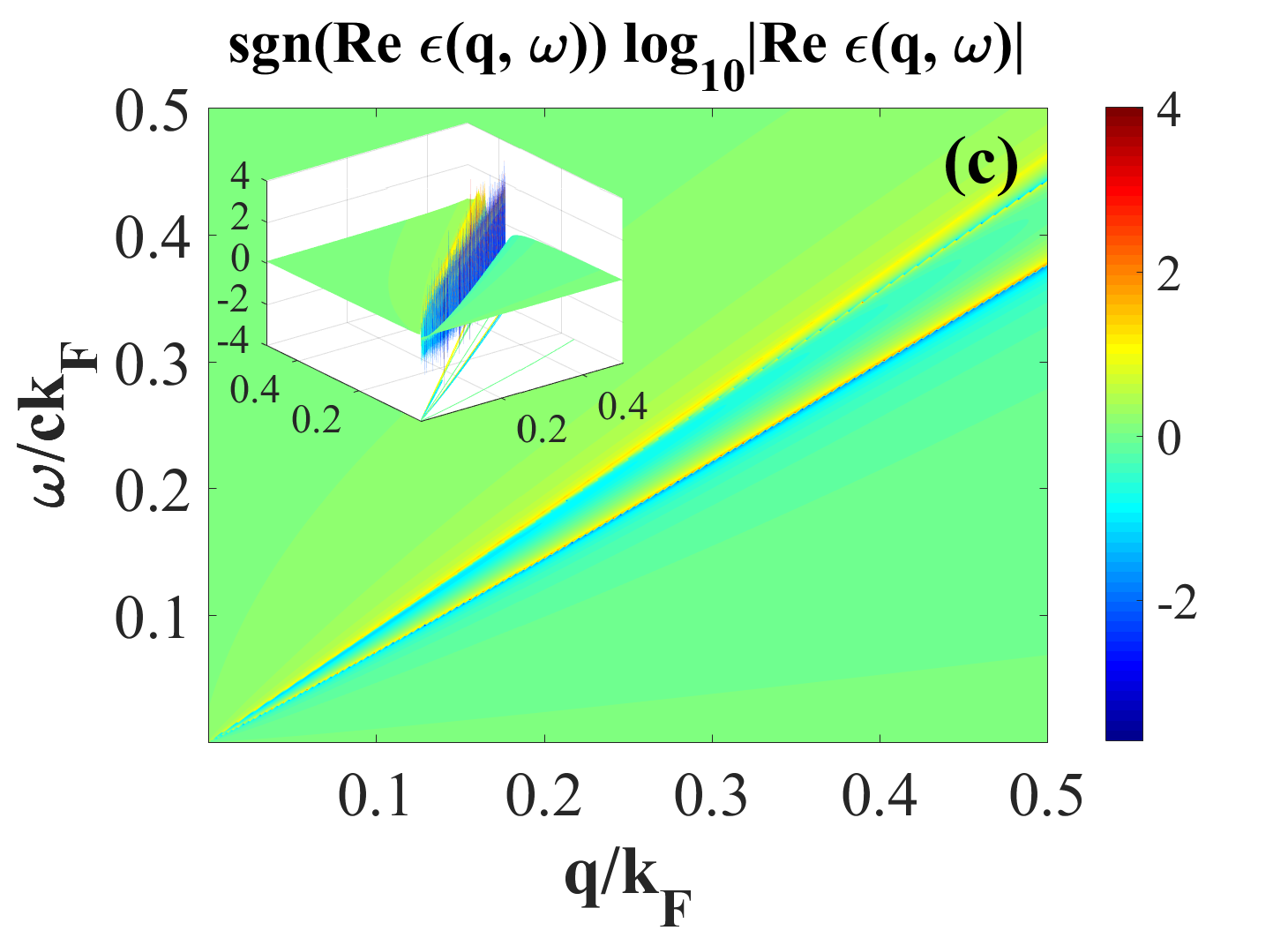}
\includegraphics[width=0.49\columnwidth]{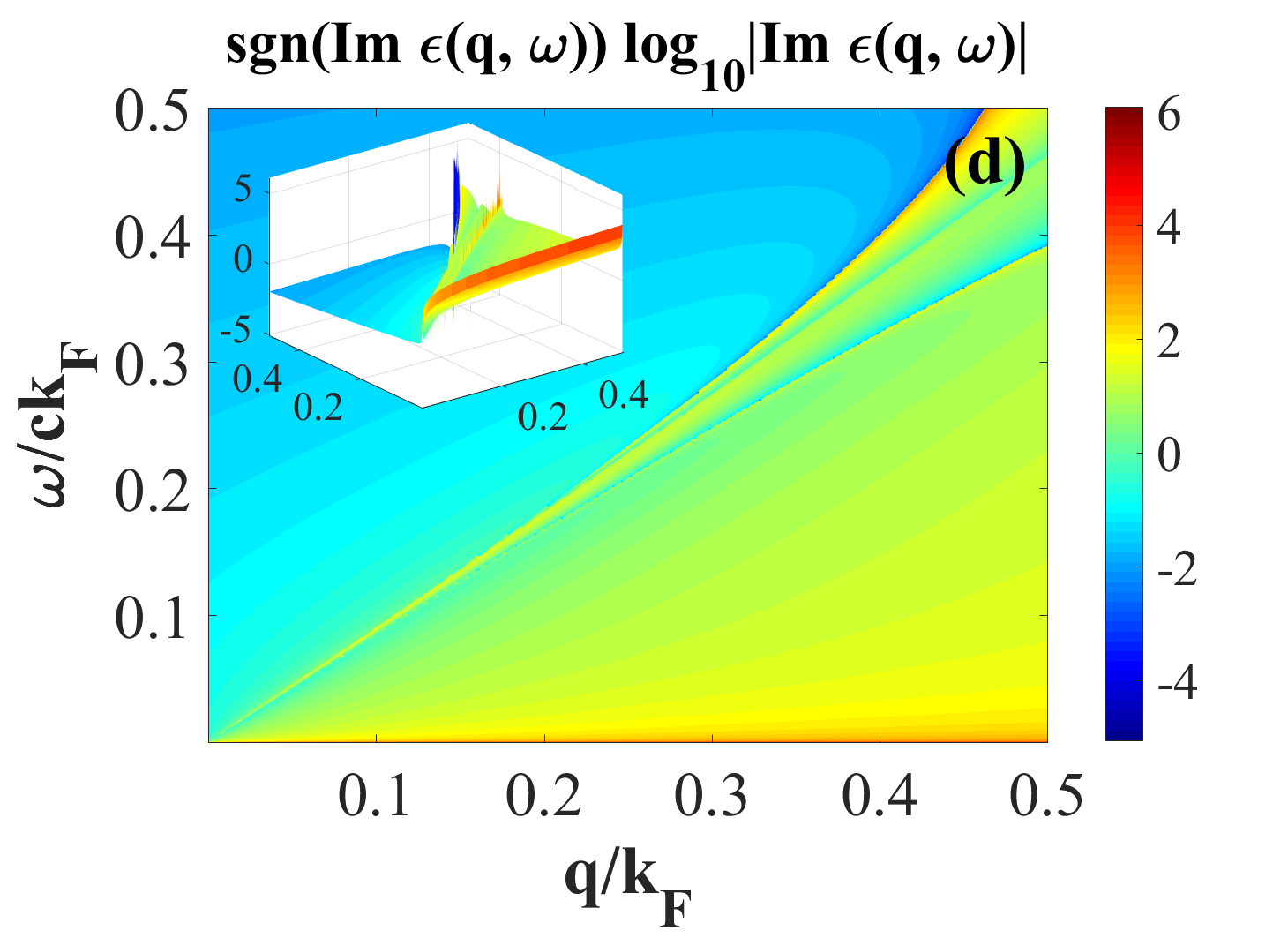}
\caption{The same as in Figure \ref{epsilon1D_T=0-001} but for higher temperature $T/ck_F=0.1$.} 
\label{epsilon1D_T=0-1}
\end{figure}

The real and imaginary part of the dielectric function for zero and nonzero ($T/ck_F=0.1$) temperatures are displayed in Figures \ref{epsilon1D_T=0-001} and \ref{epsilon1D_T=0-1}, respectively. Regardless the value of temperature, the characteristic feature of these dependencies is the oscillation behavior with the pronounced spikes of $\epsilon(q,\omega)$ in the vicinity of small values of $q$ and $\omega$. We emphasize it by means of surface plots in insets to Figures \ref{epsilon1D_T=0-001} and \ref{epsilon1D_T=0-1}.
Conceptually, increasing the temperature does not significantly change the behavior of the dielectric function. The only difference visible to the naked eye from Figures  \ref{epsilon1D_T=0-001} and \ref{epsilon1D_T=0-1}  is the reduction in the amplitude of oscillations.

\begin{figure}
\includegraphics[width=0.49\columnwidth]{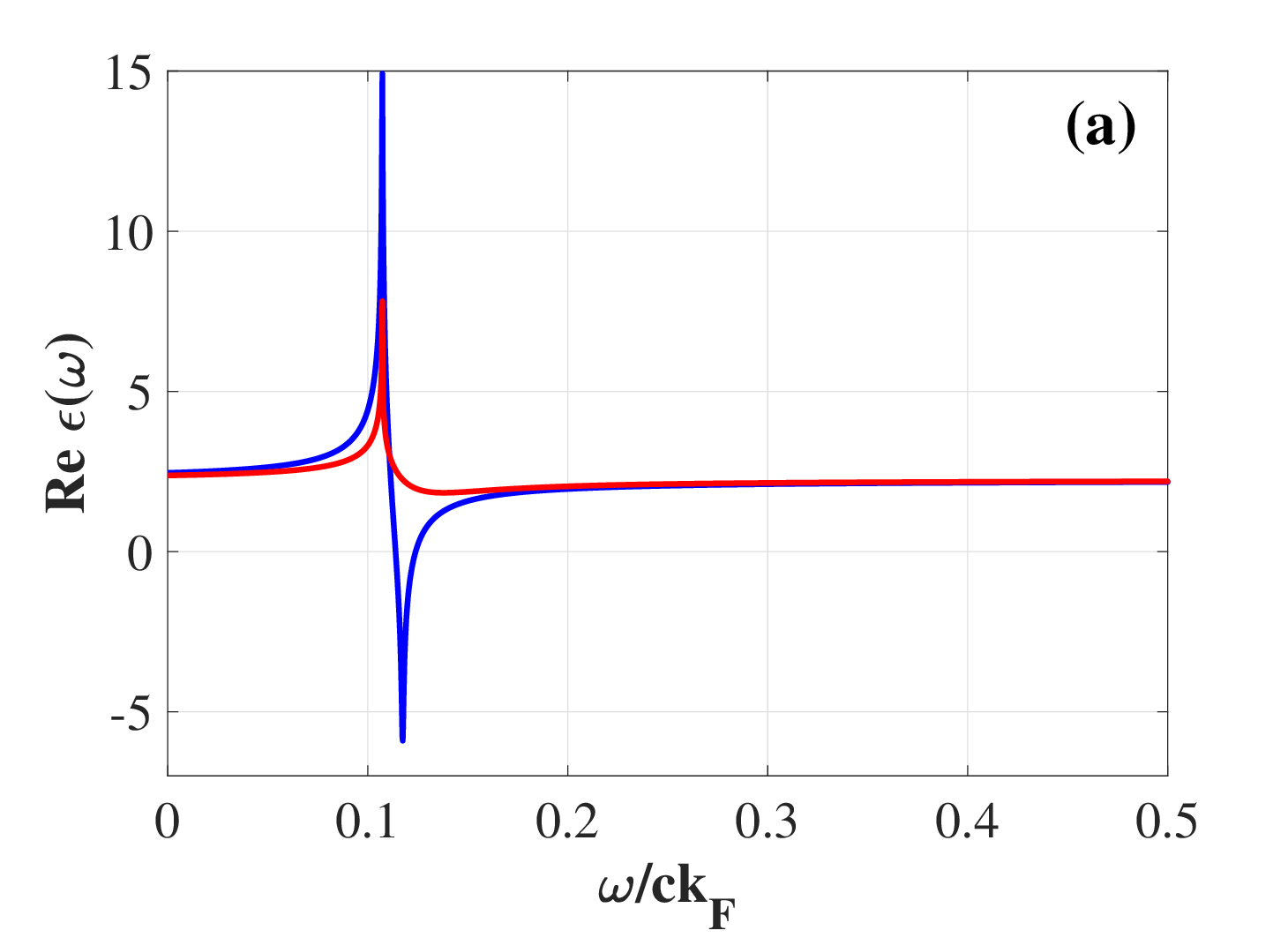}
\includegraphics[width=0.49\columnwidth]{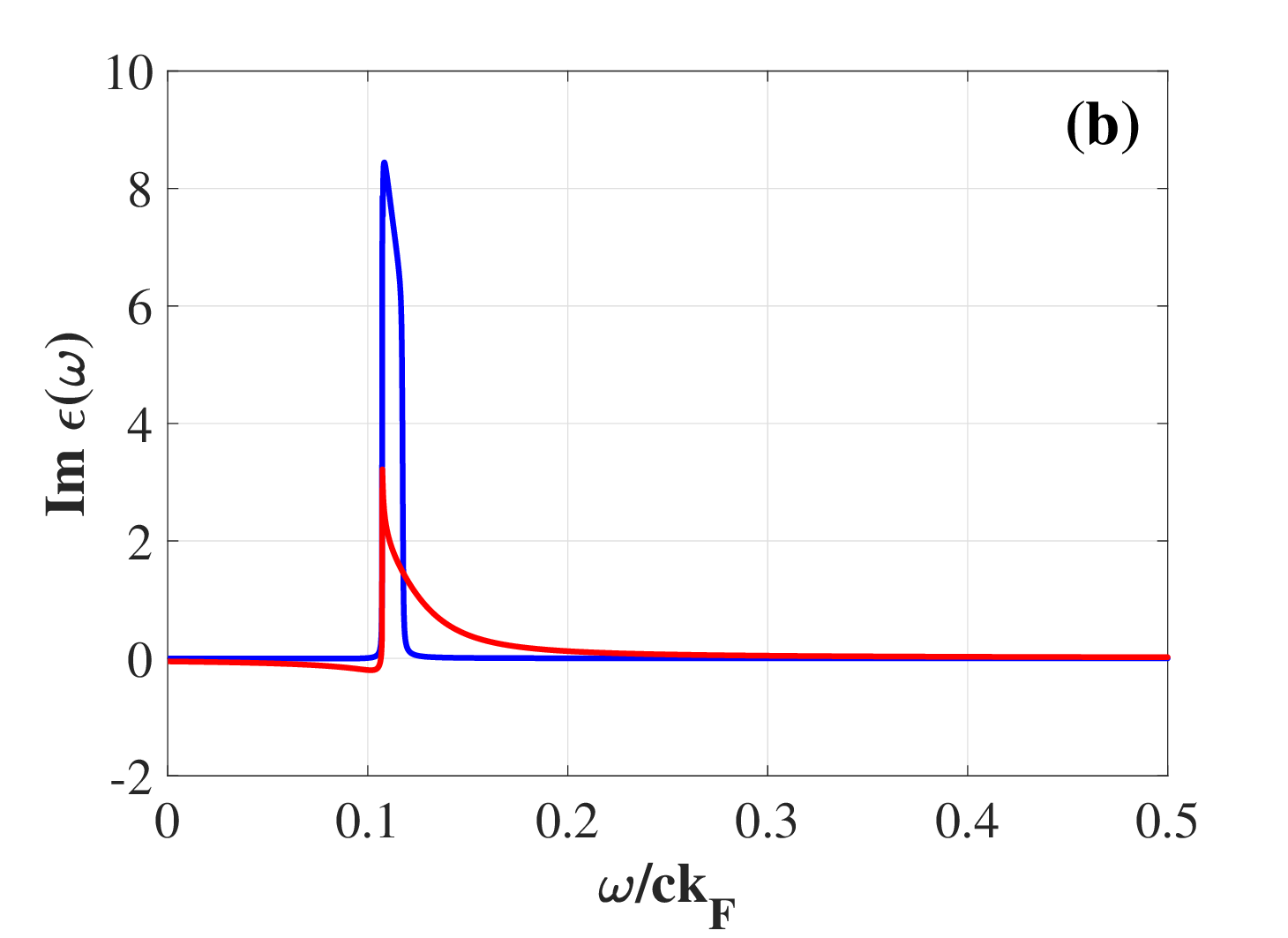}
\includegraphics[width=0.49\columnwidth]{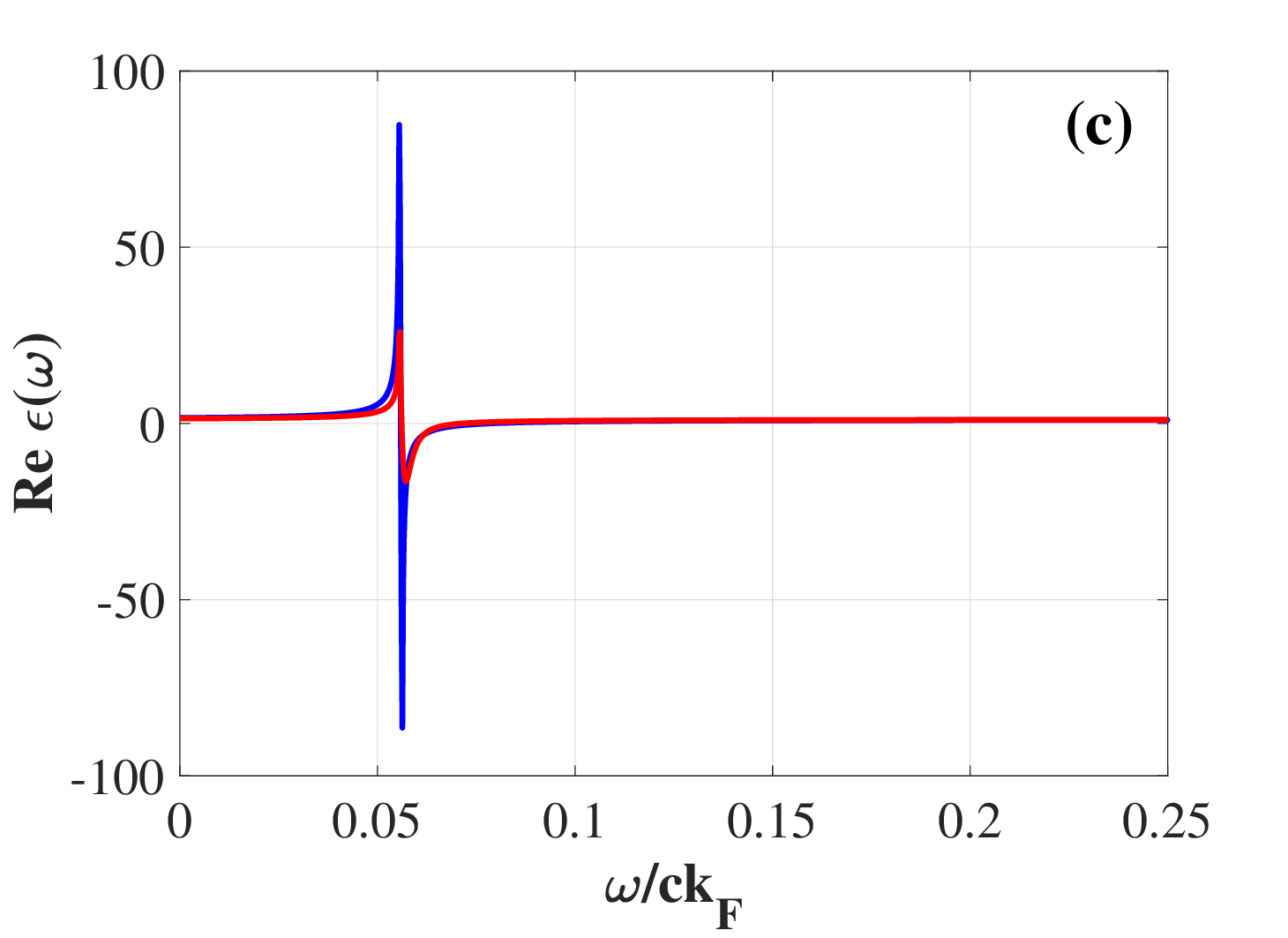}
\includegraphics[width=0.49\columnwidth]{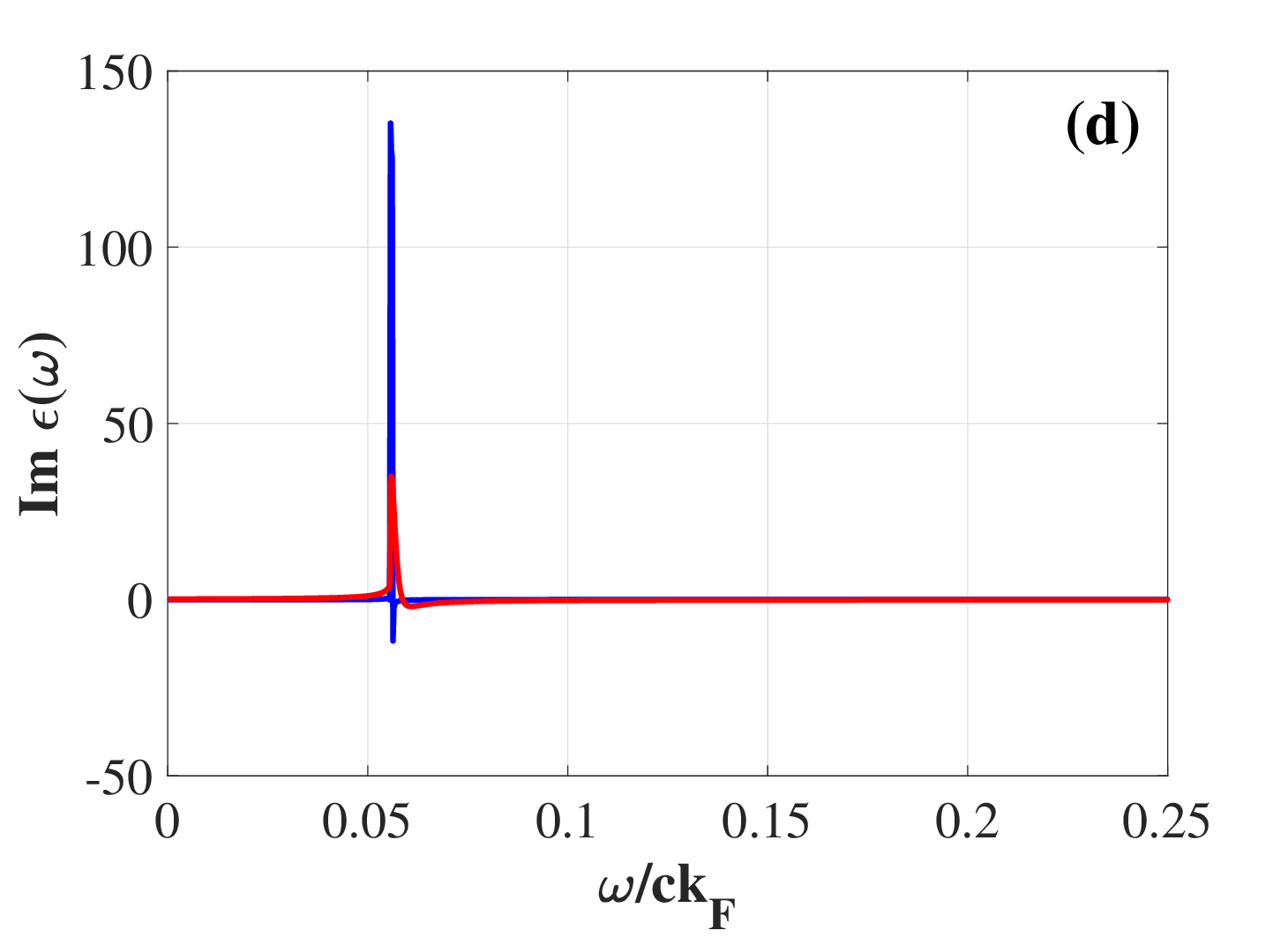}
\caption{The long-wavelength limit behavior of the the real (a, c) and imaginary (b, d) part of the dielectric finction $\epsilon(q,\omega)$ with $q=0.05k_F$ as a function of $\omega$ for a 1D gapped momentum material with the parameter $\Gamma/ck_F=4$ (a, b) and $\Gamma/ck_F=1$ (c, d) and for $T=0$ (blue line) and $T/ck_F=0.1$ (red line).} 
\label{epsilon1D_long_wave}
\end{figure}

To elucidate the change of the sign and the enhancement of the real and imaginary part of the dielectric function in the long wavelength limit $q \to 0$  we calculate separately ${\mathop{\rm Re}\nolimits} \ \epsilon (q/k_F=0.05,\omega )$ and ${\mathop{\rm Im}\nolimits} \ \epsilon (q/k_F=0.05,\omega )$ frequency dependencies. Figure \ref{epsilon1D_long_wave} demonstrates the results of corresponding calculations, which confirm our conclusions about the behaviour of $\epsilon(q,\omega)$. 

\subsubsection{Plasmon modes.}
\begin{figure}
\includegraphics[width=0.49\columnwidth]{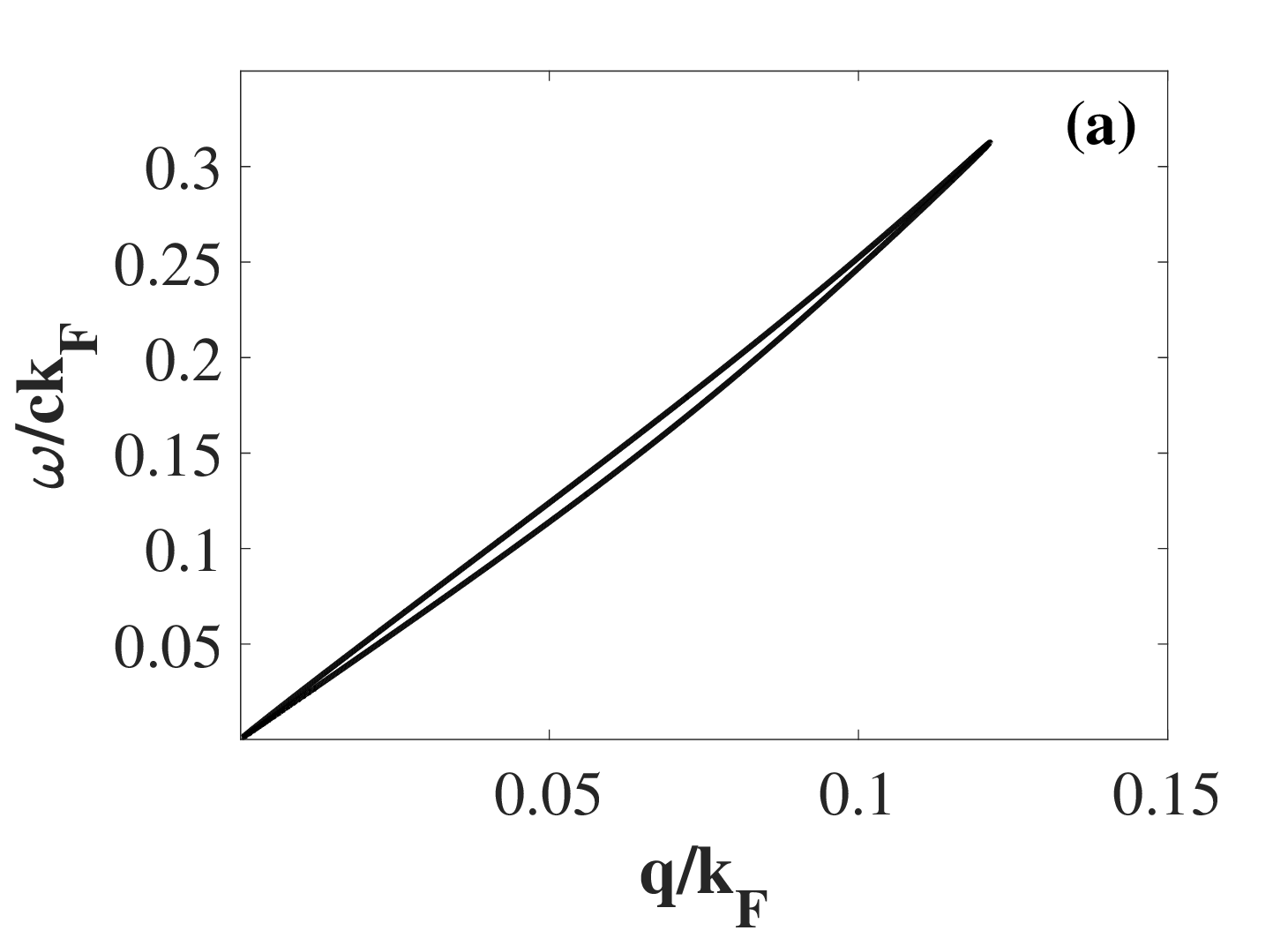}
\includegraphics[width=0.49\columnwidth]{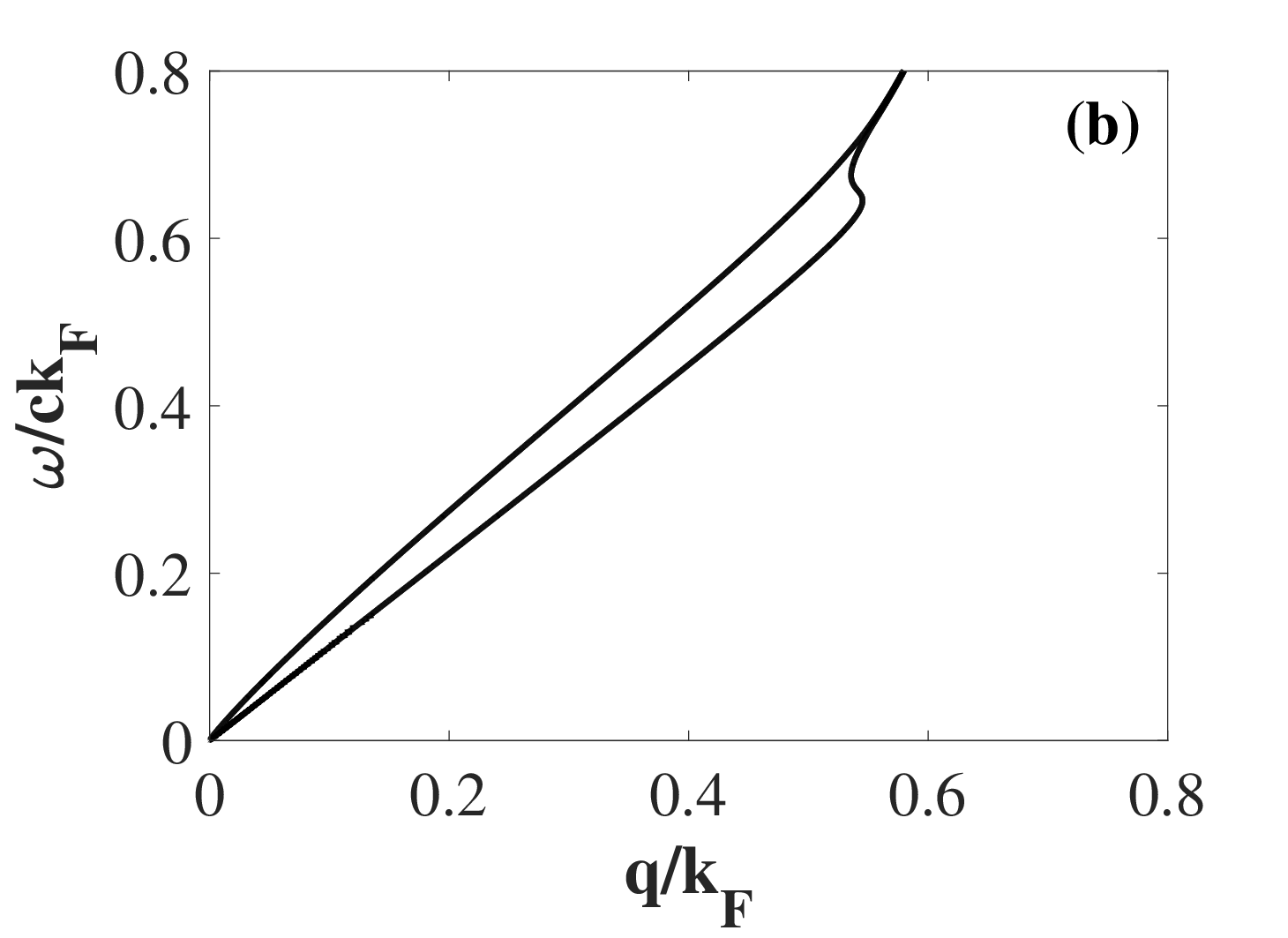}
\caption{Plasmon dispersion relations for a 1D gapped momentum material with $\Gamma/ck_F=4$ (a) and $\Gamma/ck_F=1$ (b) at $T=0$.} 
\label{plasmon1D_T=0-001}
\end{figure}

To determine the plasmon dispersion $\omega^{1D}_{pl}(q)$ we proceed to the solution of Eq. (\ref{plasmon_general}). According to the numerical results presented in Figure \ref{plasmon1D_T=0-001} the plasmon dispersion acquires a notable feature: the splitting of the plasmon dispersion curve into two different branches when $q$ differs from zero with their further confluence in the region of high momenta. After that one can observe their merging into the linear dependence again with further increasing of $q$. Interestingly, that similar plasmon splitting and the broadening of the plasmon line  was found also in the two-dimensional electron gas with the spin-orbit coupling \cite{Pletyukhov}. The same phenomenon was predicted also in graphene \cite{Mikhailov} with its linear dispersion relation.  

Based on the analytical ``receipt'' provided in the previous section we derive $\omega^{1D}_{pl}(q)$ dependence analytically in the long-wavelength limit at $T=0$. Restricting our consideration by the first two terms in the expansion of Eq. (\ref{plasmon_approx_eq}) and performing the straightforward integration by means of the Sokhotski–Plemelj theorem we obtain
\begin{widetext}
\begin{equation}
\label{plasmon_approx_1D}
\omega^{1D}_{pl}(q)  = \frac{{{e^2}{K_0}\left( {qa} \right)}}{{2\pi }}\left( {1 + \sqrt {1 + \frac{{4\pi c}}{{{e^2}{K_0}\left( {qa} \right)}}\frac{{\sqrt {2{\mkern 1mu} {\kern 1pt} \sqrt {{\mu ^4} + {\mu ^2}{\Gamma ^2}}  - 2{\mkern 1mu} {\kern 1pt} {\mu ^2}} \Gamma  + 2{\mkern 1mu} {\kern 1pt} \sqrt {2{\mkern 1mu} {\kern 1pt} \sqrt {{\mu ^4} + {\mu ^2}{\Gamma ^2}}  + 2{\mkern 1mu} {\kern 1pt} {\mu ^2}} \mu }}{{{\Gamma ^2} + 4{\mkern 1mu} {\kern 1pt} {\mu ^2}}}} } \right)q.
\end{equation}
\end{widetext}
One can verify that this expression agrees with that one for plasmon dispersion in Dirac gapped materials \cite{Thakur}. To this end we recall that gapped momentum media can be interpreted in some ways as Dirac gapped systems with the complex value of the energy gap. Therefore, the coincidence of both dependencies $\omega_{pl}(q)$ can be considered as the proof of the reliability of our calculations. 
  
\subsection{2D gapped momentum material}

\subsubsection{The dielectric function}

Having studied the 1D gapped momentum medium, we now focus on its 2D counterpart. For $T=0$ the polarization function Eq. (\ref{Polarization_general}) admits some analytical simplifications resulting in the integration of the complete elliptic integrals of the first and the third kind (see Appendix \ref{sec:B} for details).  In the limiting case of zero temperature one can perform the angular integration in Eq. ({\ref{Polarization_general}). This analytical simplification allows to verify our further numerical procedure for calculation of the dielectric function in more general case of nonzero temperature.
\begin{figure}
\includegraphics[width=0.49\columnwidth]{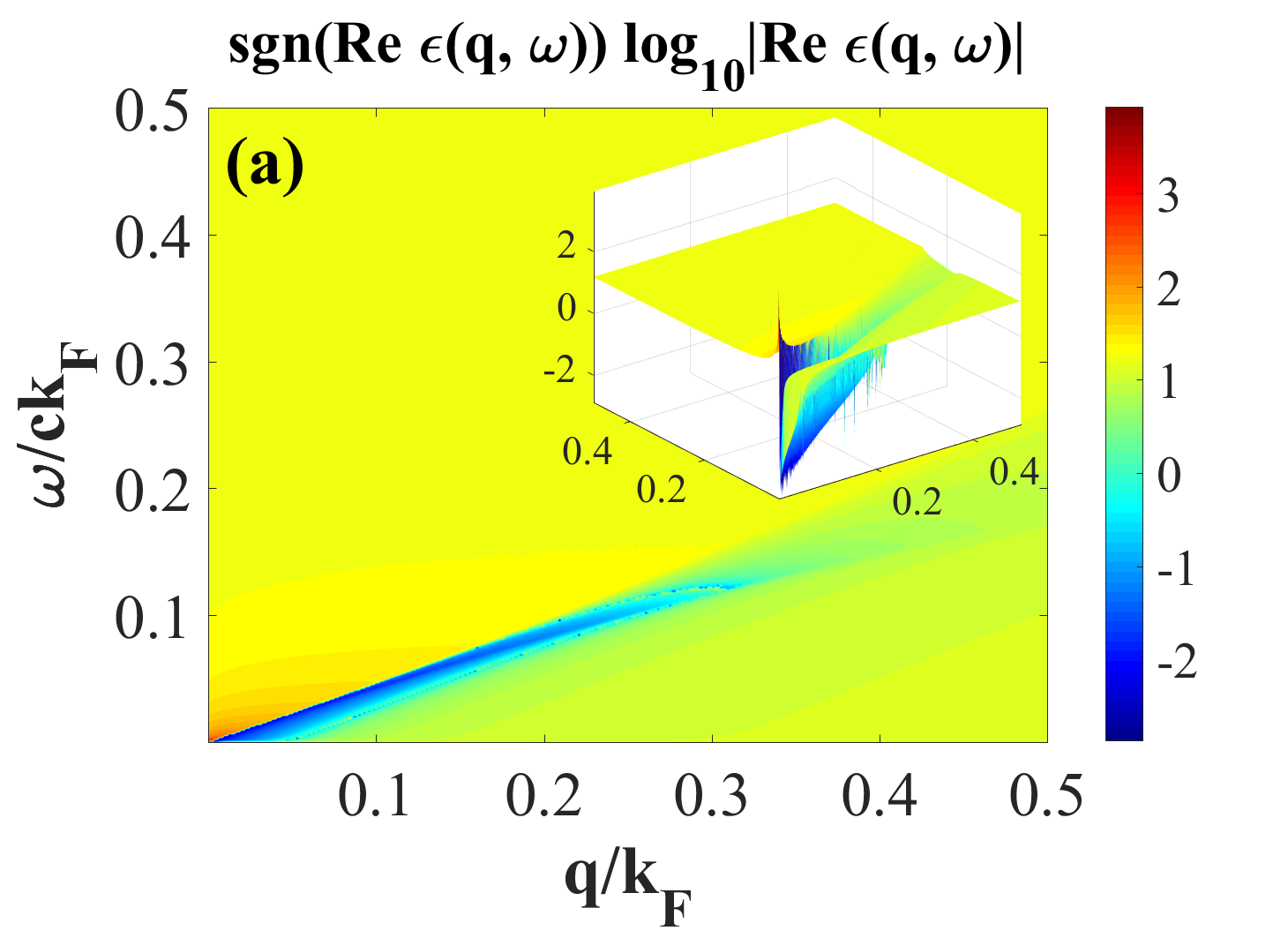}
\includegraphics[width=0.49\columnwidth]{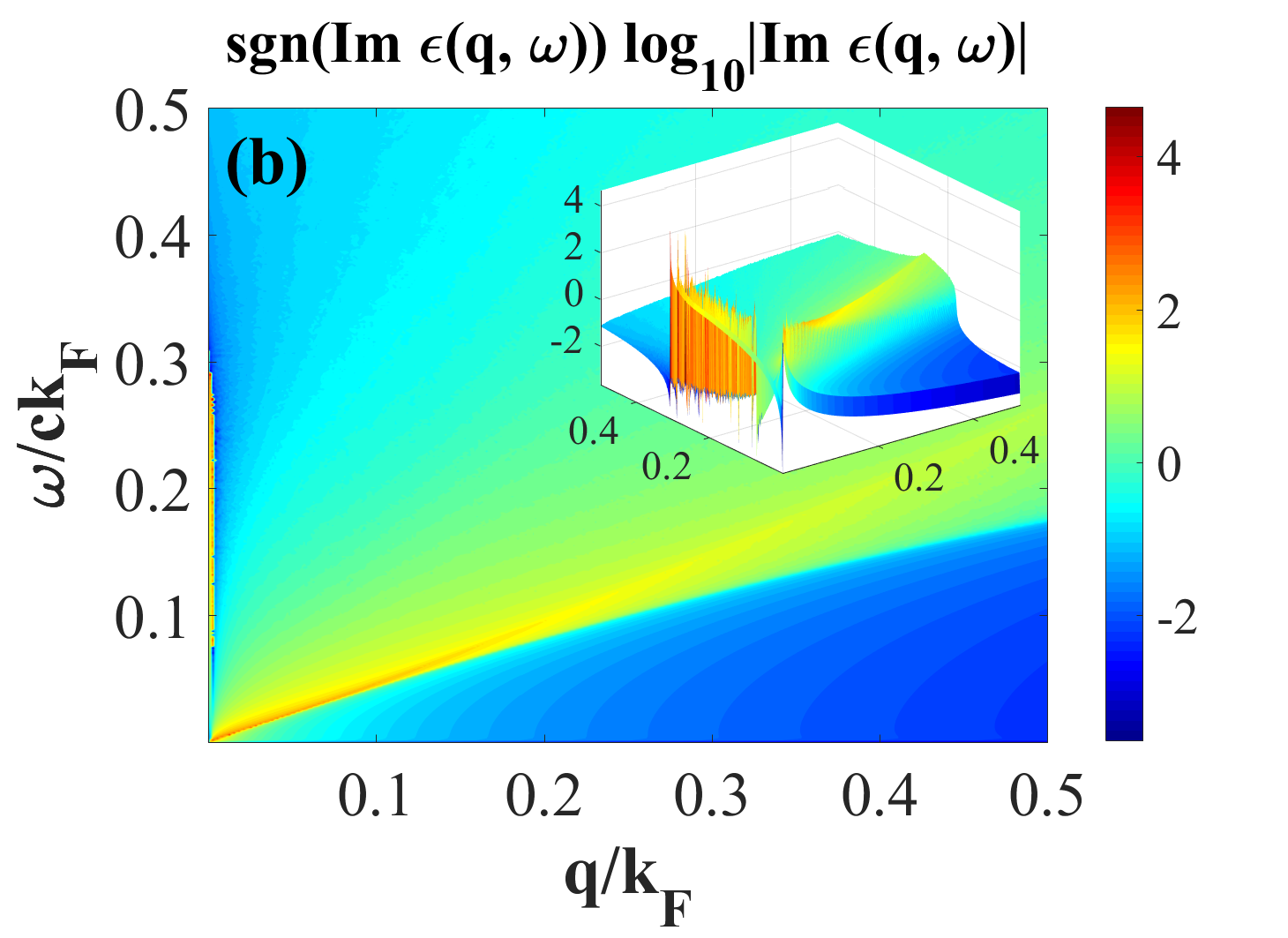}
\includegraphics[width=0.49\columnwidth]{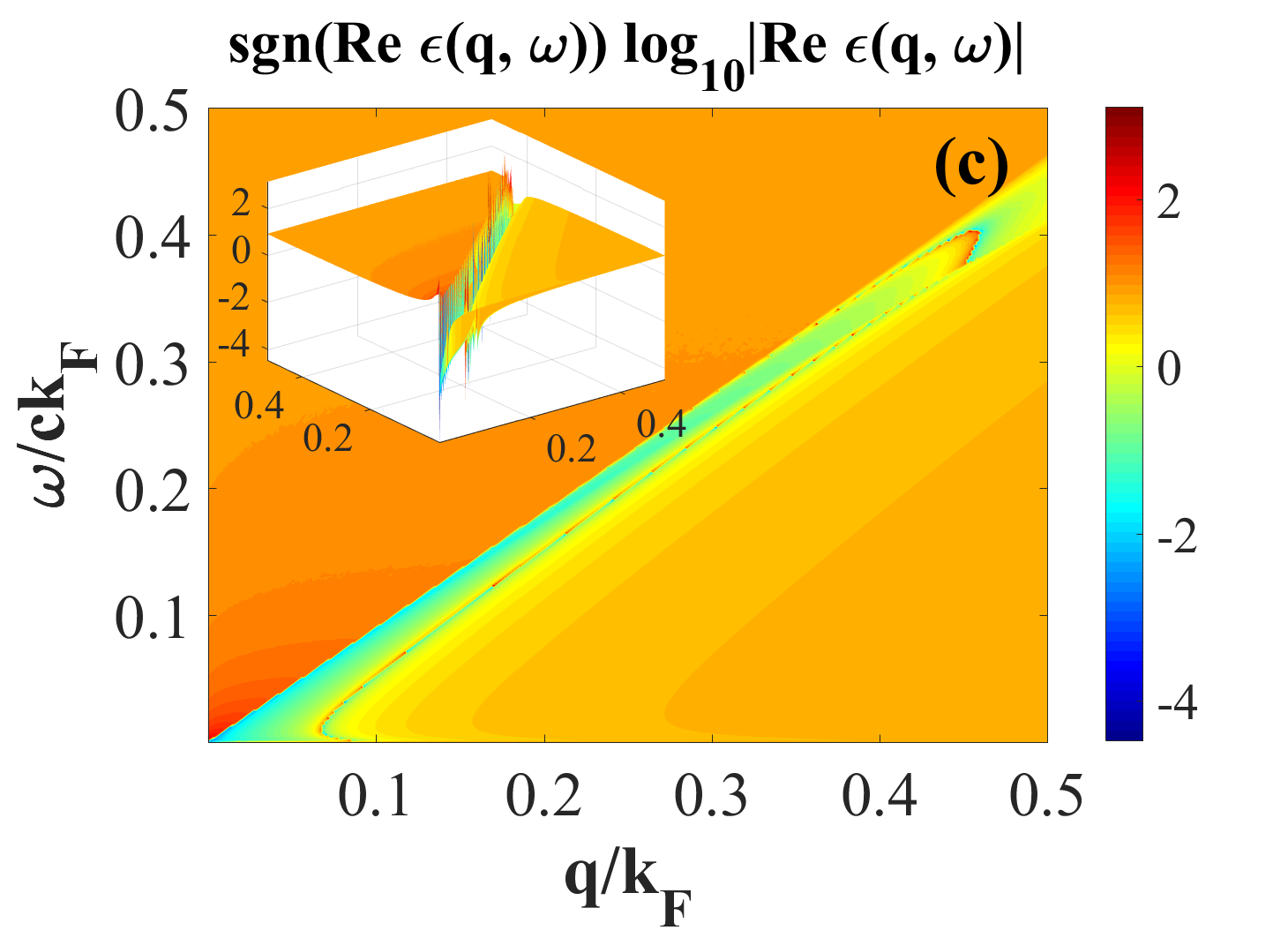}
\includegraphics[width=0.49\columnwidth]{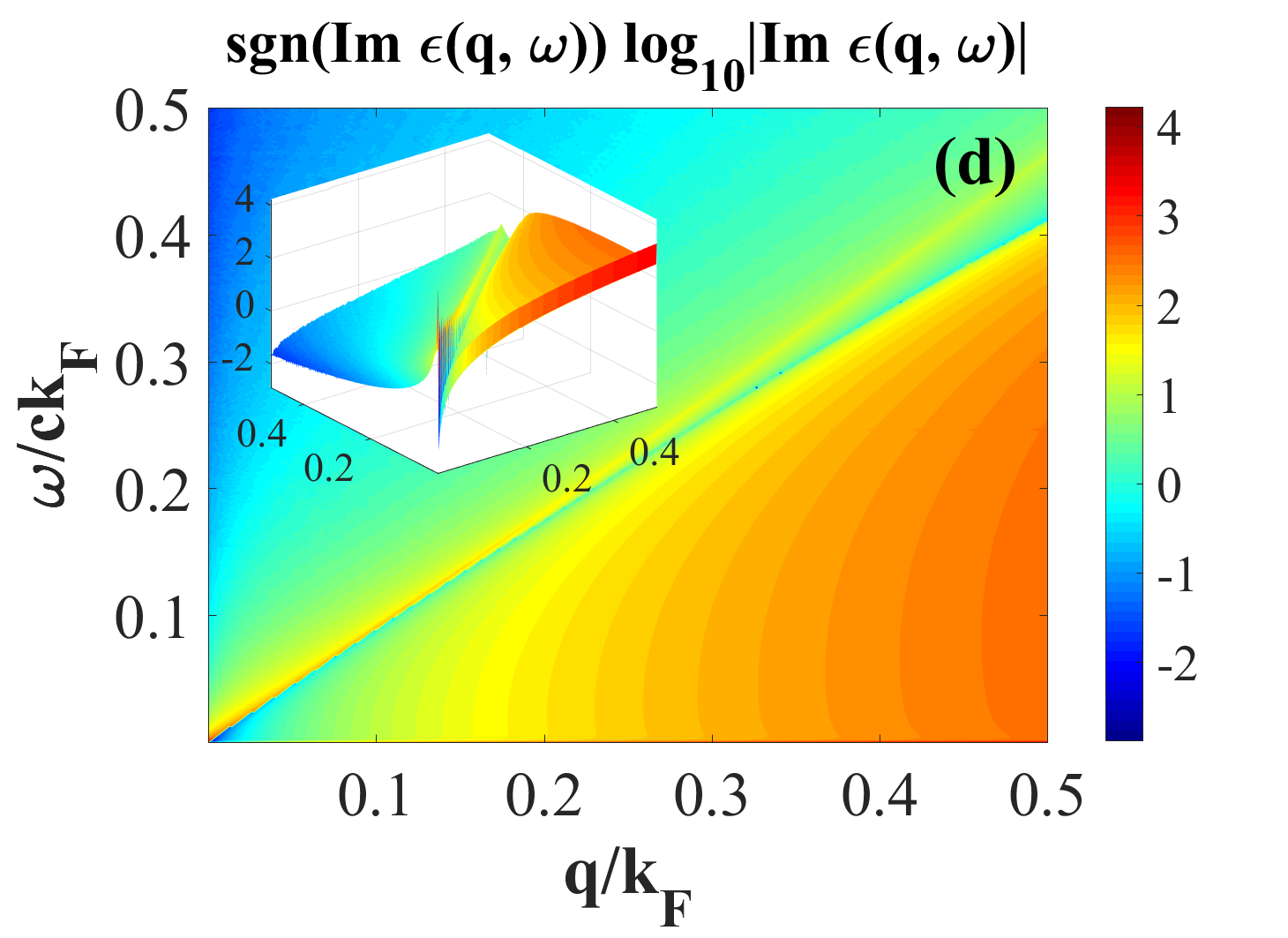}
\caption{The real (a, c) and imaginary (b, d) part of the dielectric finction $\epsilon(q,\omega)$ in $q$-$\omega$ plane for a 2D gapped momentum material with the parameter $\Gamma/ck_F=4$ (a,b) and $\Gamma/ck_F=1$ (a,b) and for the for the temperature $T=0$. Insets show three-dimensional surface plot of $\epsilon(q,\omega)$ to illustrate additionally the behavior of the dielectric function, which is not clear visible in contour plots in the region of small $q$ and $\omega$.} 
\label{epsilon2D_T=0-001}
\end{figure}
Performing numerical integration of Eq. (\ref{epsilon_general}) in 2D case for $T=0$ we obtain the real and imaginary parts of $\epsilon(q,\omega)$ as a function of the wave vector $q$ and the frequency $\omega$ which are visualized in Figure \ref{epsilon2D_T=0-001}. Similarly to the 1D gapped system we observe oscillations of the dielectric function with the growing amplitude in the domain of small values of $q,\omega$, but beyond this region the oscillations decay more rapidly in comparison with the 1D case (see insets of Figure \ref{epsilon2D_T=0-001}). 

 At nonzero temperature ($T/ck_F=0.1$) the 2D dielectric function qualitatively follows its behavior in the 1D case (a decrease of the amplitude of $\epsilon(q,\omega)$). In order not to overload the paper with an excessive number of figures, we do not provide the details of corresponding results. Instead, we study more carefully the frequency dependence of $\epsilon(q,\omega)$ in the long-wavelength limit, where these oscillations are observed. One can see in Figure \ref{epsilon2D_long_wave} the frequency dependence of the real and the imaginary parts of $\epsilon(q=0.05\,k_F,\omega)$ for $\Gamma/ck_F=4$ and $\Gamma/ck_F=1$ and for zero and nonzero temperatures.
\begin{figure}
\includegraphics[width=0.49\columnwidth]{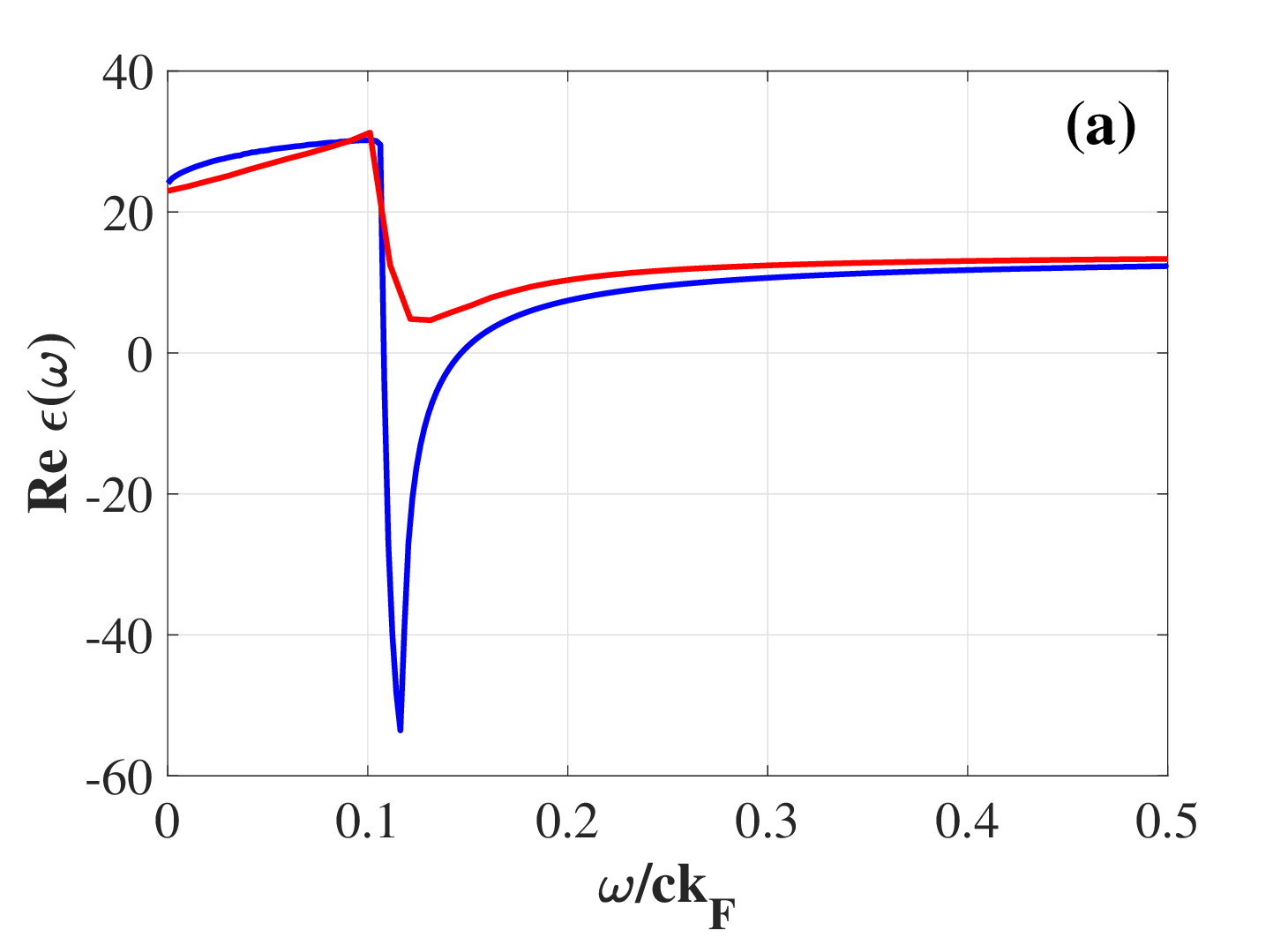}
\includegraphics[width=0.49\columnwidth]{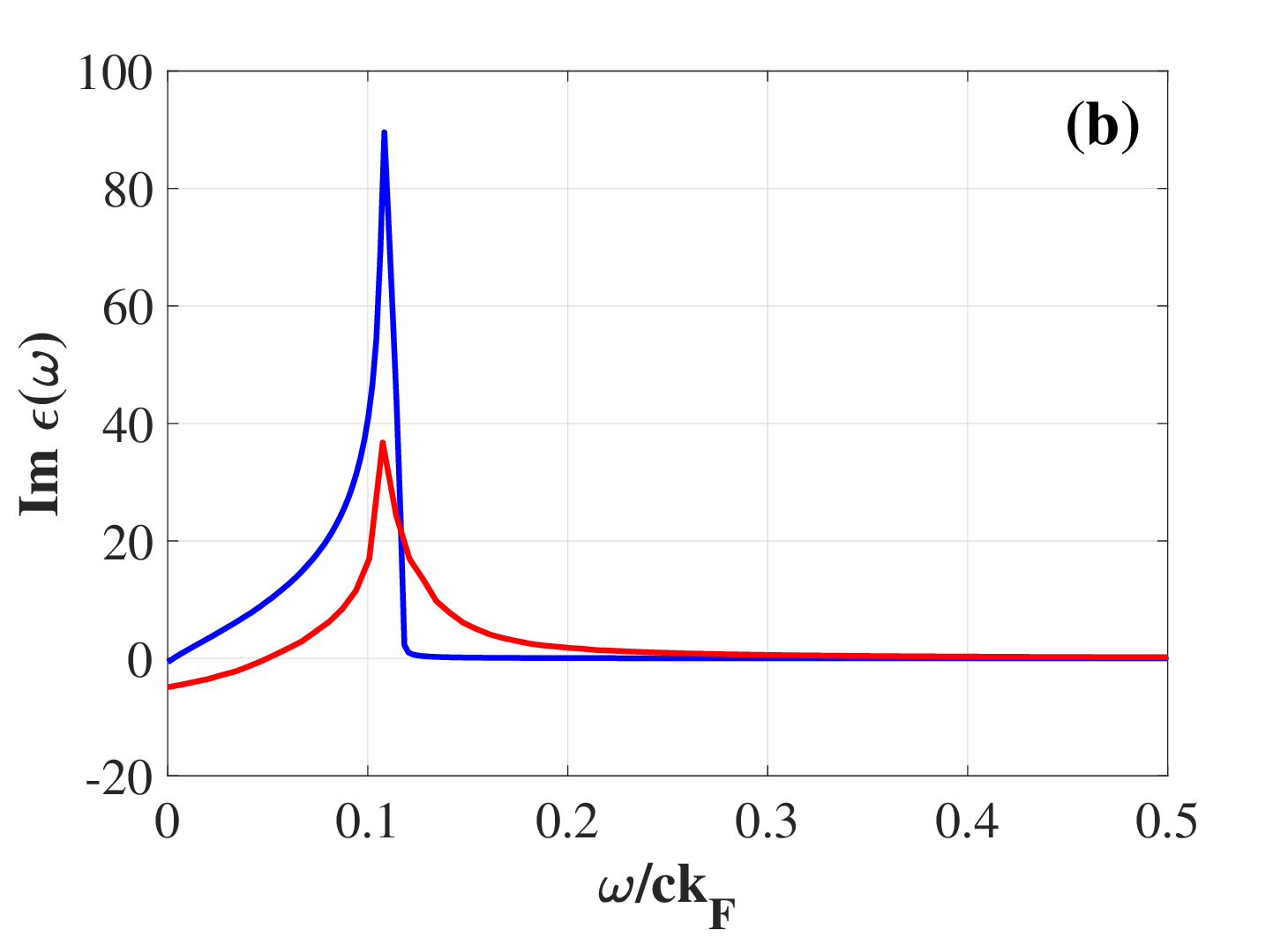}
\includegraphics[width=0.49\columnwidth]{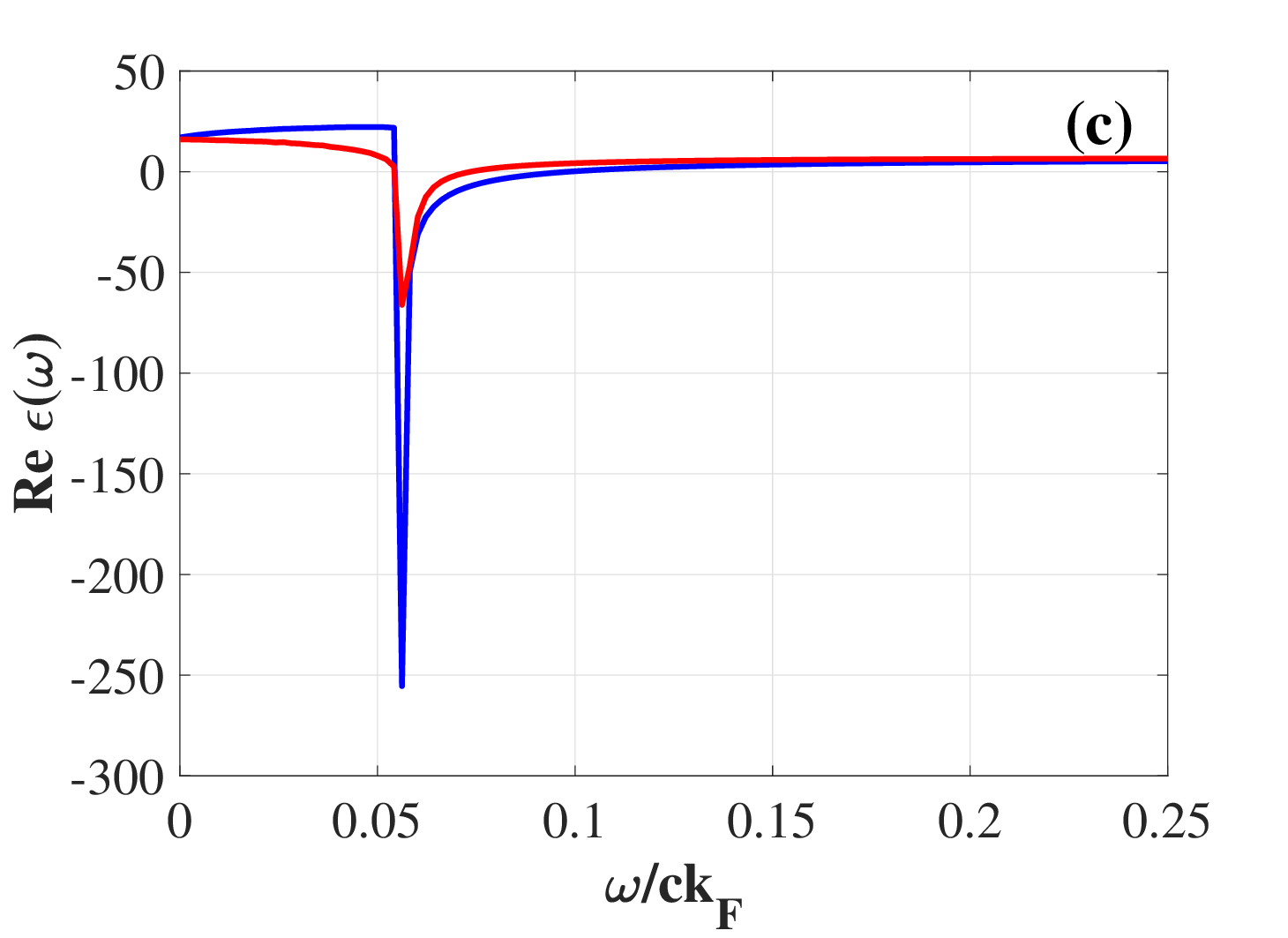}
\includegraphics[width=0.49\columnwidth]{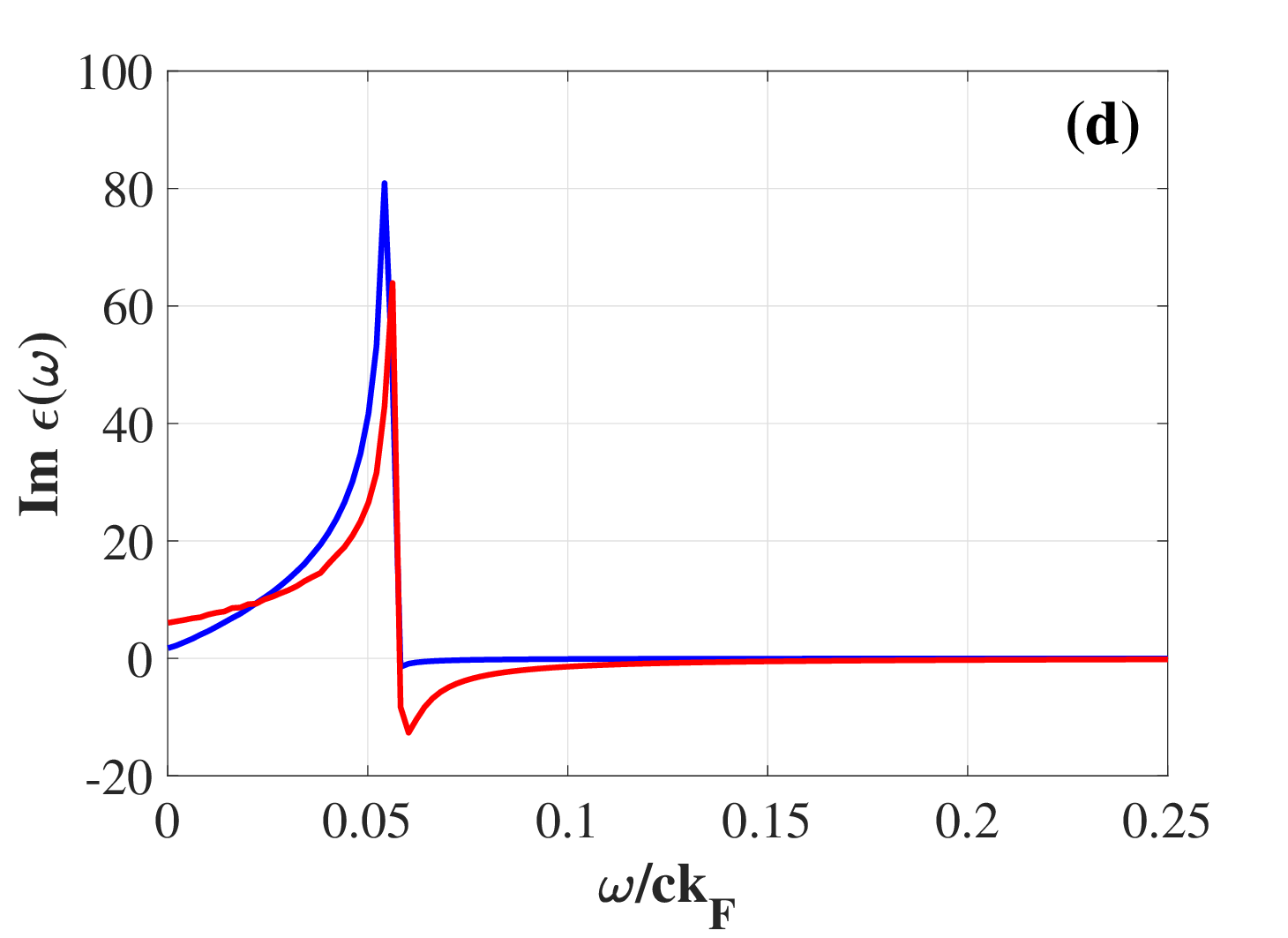}
\caption{The long-wavelength limit behavior of the the real (a, c) and imaginary (b, d) part of the dielectric function $\epsilon(q,\omega)$ with $q=0.05k_F$ as a function of $\omega$ for a 2D gapped momentum material with the parameter $\Gamma/ck_F=4$ (a, b) and $\Gamma/ck_F=1$ (c, d) and for $T=0$ (blue line) and $T/ck_F=0.1$ (red line).} 
\label{epsilon2D_long_wave}
\end{figure}
Figure \ref{epsilon2D_long_wave} clearly shows the temperature suppression of the dielectric function, confirming our conclusions about the evolution of $\epsilon(q,\omega)$, when $T$ grows.

\subsubsection{Plasmon modes}
Knowing the behaviour of the dielectric function we can retrieve  information about the plasmon dispersion relation. According to our numerical results the latter undergoes the phenomenon of splitting of plasmon modes (see Figure \ref{plasmon2D_T=0-001}) analogous to the 1D case.
\begin{figure}
\includegraphics[width=0.49\columnwidth]{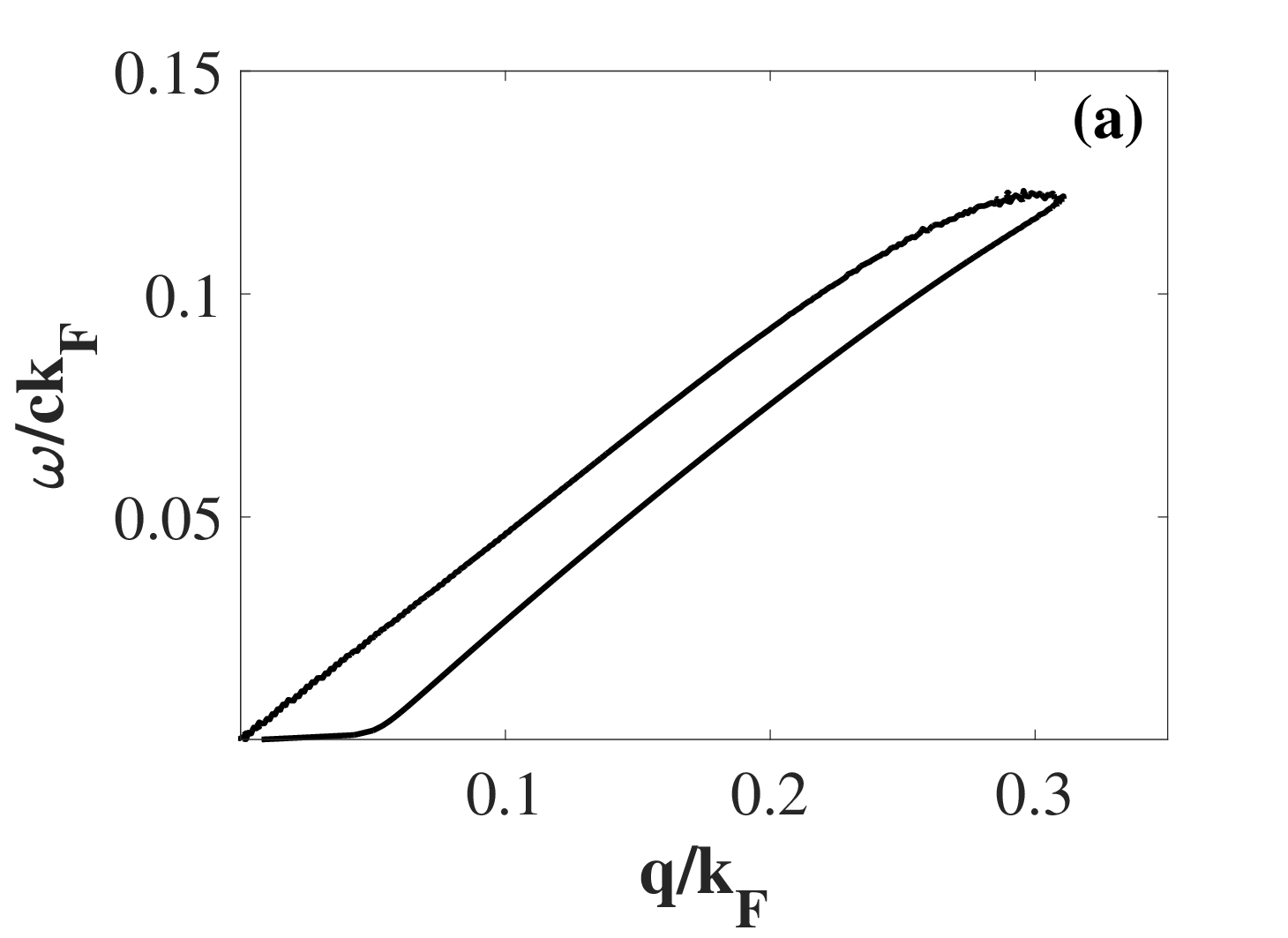}
\includegraphics[width=0.49\columnwidth]{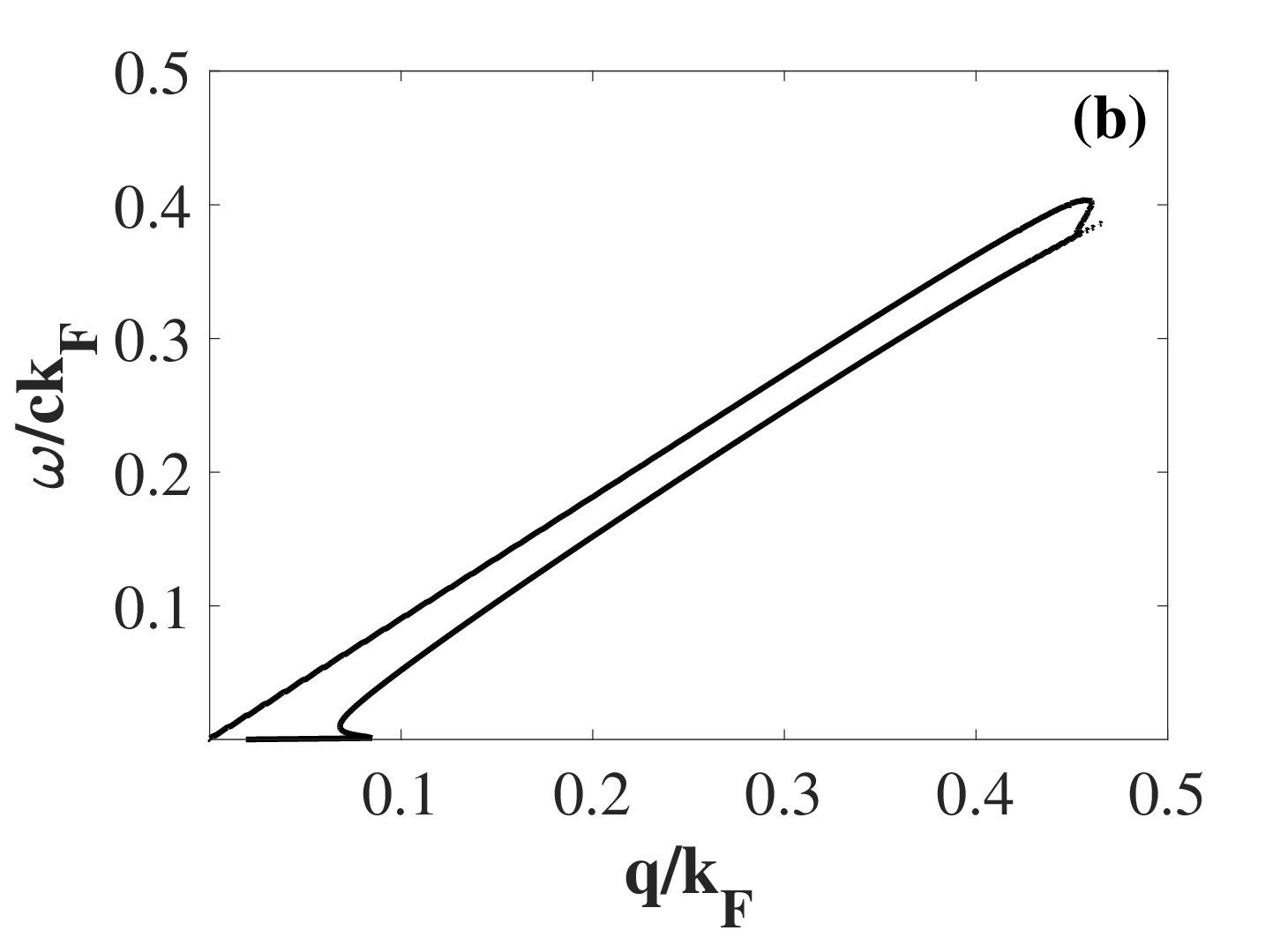}
\caption{Plasmon dispersion relations for a 2D gapped momentum material with $\Gamma/ck_F=4$ (a) and $\Gamma/ck_F=1$ (b) at low temperature $T/ck_F=0$.} 
\label{plasmon2D_T=0-001}
\end{figure}
As already discussed in the previous section, the behavior of $\omega^{2D}_{pl}(q)$  reminds the similar dependence for the two-dimensional electron gas with the spin-orbit coupling and graphene.

Fig. \ref{plasmon2D_T=0-001} shows that with decreasing $\Gamma$ coefficient the plots of the implicit function $\omega^{2D}_{pl}(q)$ extends on the $q$-$\omega$ plane like the analogous dependence for the 1D case (see Fig. \ref{plasmon1D_T=0-001}). Apparently, this is because the smaller is the value of the $\Gamma$ coefficient, the larger is the interval of real values for the momentum $k$ and correspondingly for the energy $E_k$ in the electronic dispersion relation given by Eq. (\ref{dispersion_general}).

Expanding as above Eq. (\ref{plasmon_approx_eq}) and performing integration in it according to the Sokhotski–Plemelj theorem we find
\begin{equation}
\label{plasmon_approx_2D}
\omega^{2D}_{pl}(q)  = e\sqrt {{\kern 1pt} \mu \frac{{2{\mu ^2} + {\Gamma ^2}}}{{4{\mkern 1mu} {\kern 1pt} {\mu ^2} + {\Gamma ^2}}}q}.
\end{equation}
As in the case of 2D systems with the parabolic dispersion, gapped and massless Dirac fermions, the plasmon frequency $\omega_{pl}(q)$ in the gapped momentum material turns to be proportional to $\sqrt{q}$ with the another density dependent prefactor only \cite{Thakur}. 

\subsection{3D gapped material}
\subsubsection{The dielectric function.}

This section is devoted to the consideration of the dielectric function behavior for the 3D gapped momentum medium. Similar to 1D and 2D cases we reduce the polarization function at $T=0$ to the expression containing the single integration over momentum modulus instead of the triple integral over the  Brillouin zone. The result of the analytical evaluation in view of its cumbersome structure is given in Appendix \ref{sec:C}. Eq. (\ref{polarization3D_exact}) serves as the litmus test of the accuracy during the direct (without analytical simplifications) numerical calculations of the dielectric function. In figure \ref{epsilon3D_T=0-001} the graphic representation of the behavior of the real and imaginary part of the dielectric function at $T=0$ resulting from the numerical calculation of Eq. (\ref{epsilon_general}) is given.
\begin{figure}
\includegraphics[width=0.49\columnwidth]{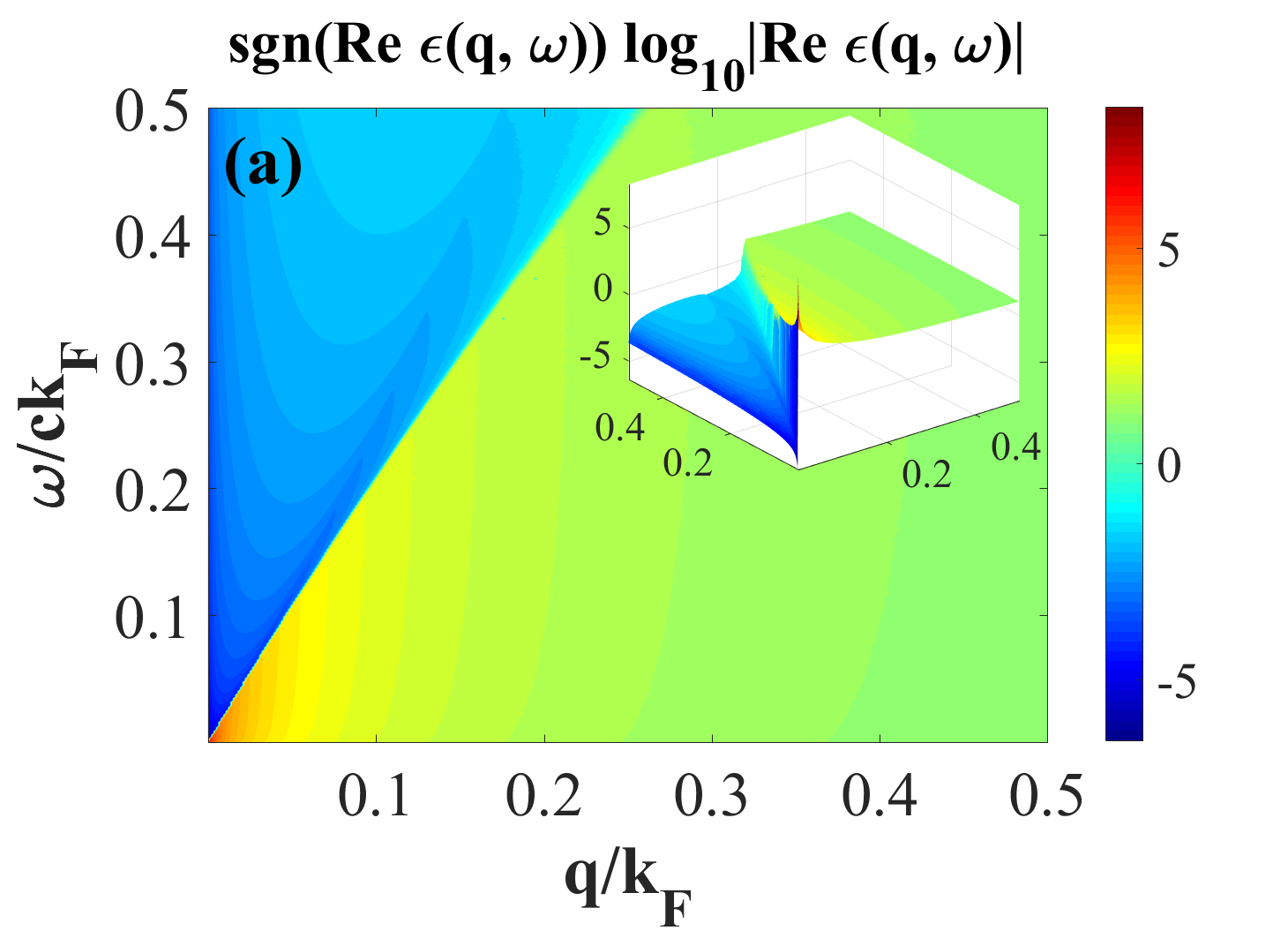}
\includegraphics[width=0.49\columnwidth]{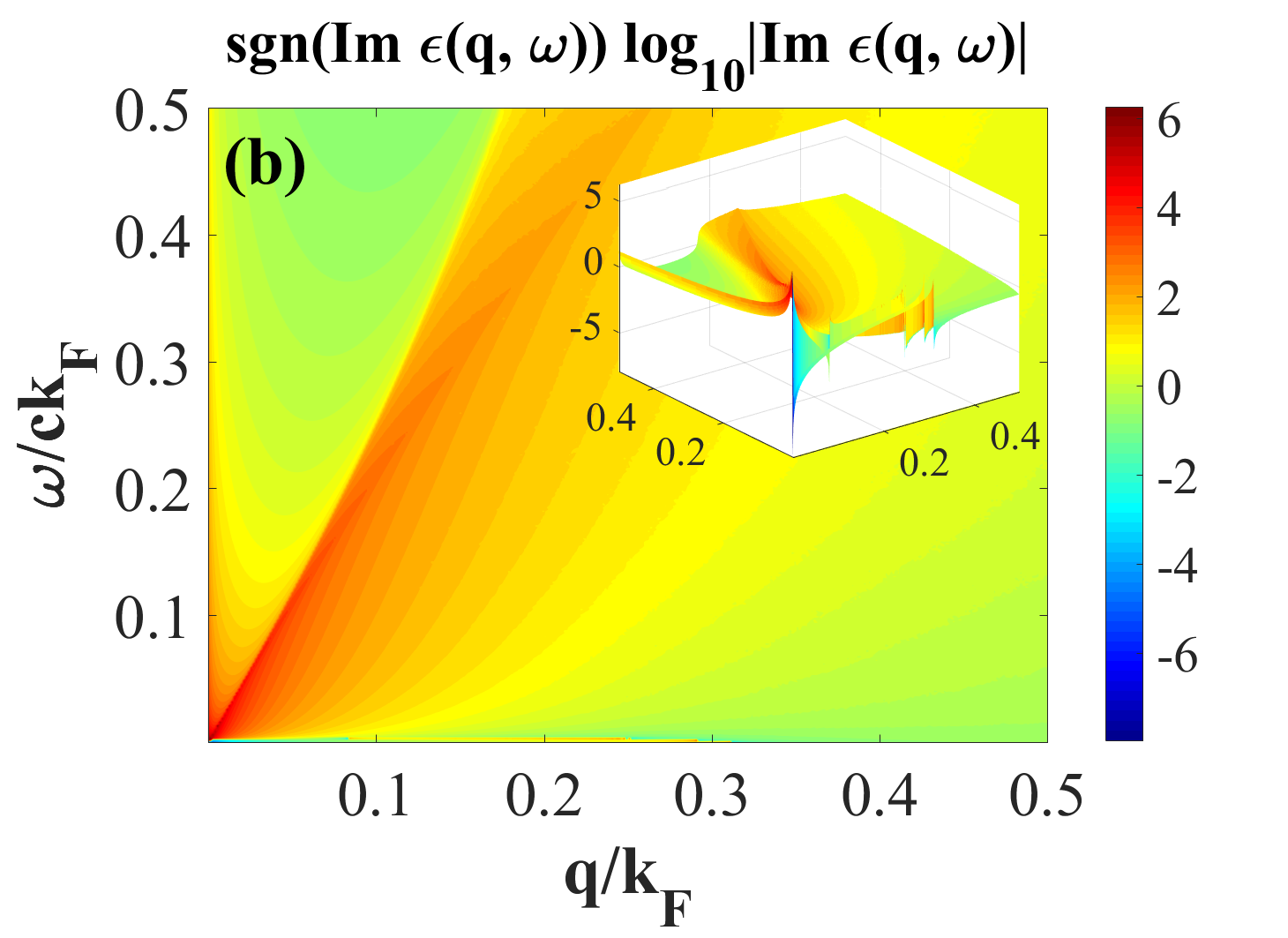}
\includegraphics[width=0.49\columnwidth]{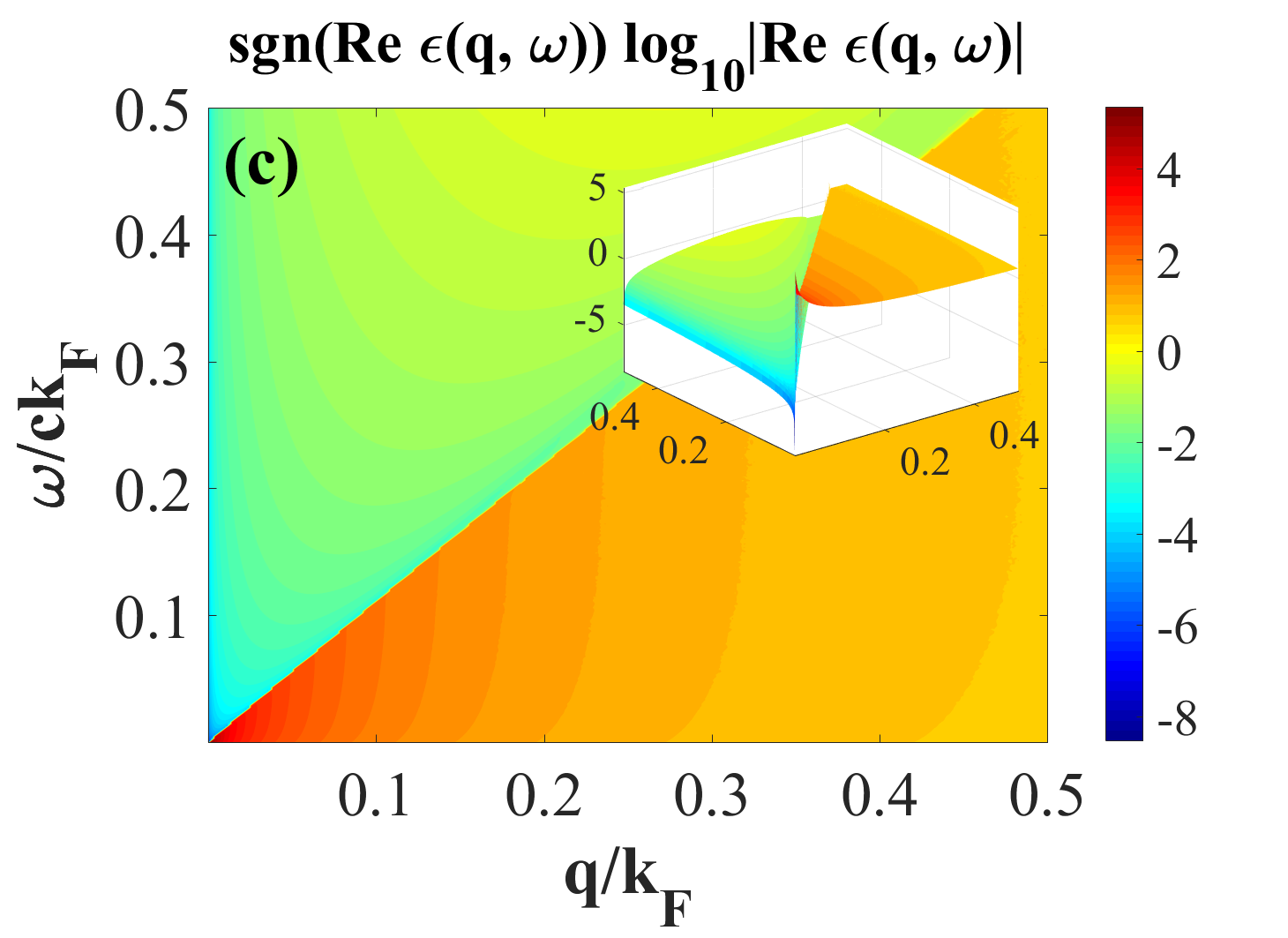}
\includegraphics[width=0.49\columnwidth]{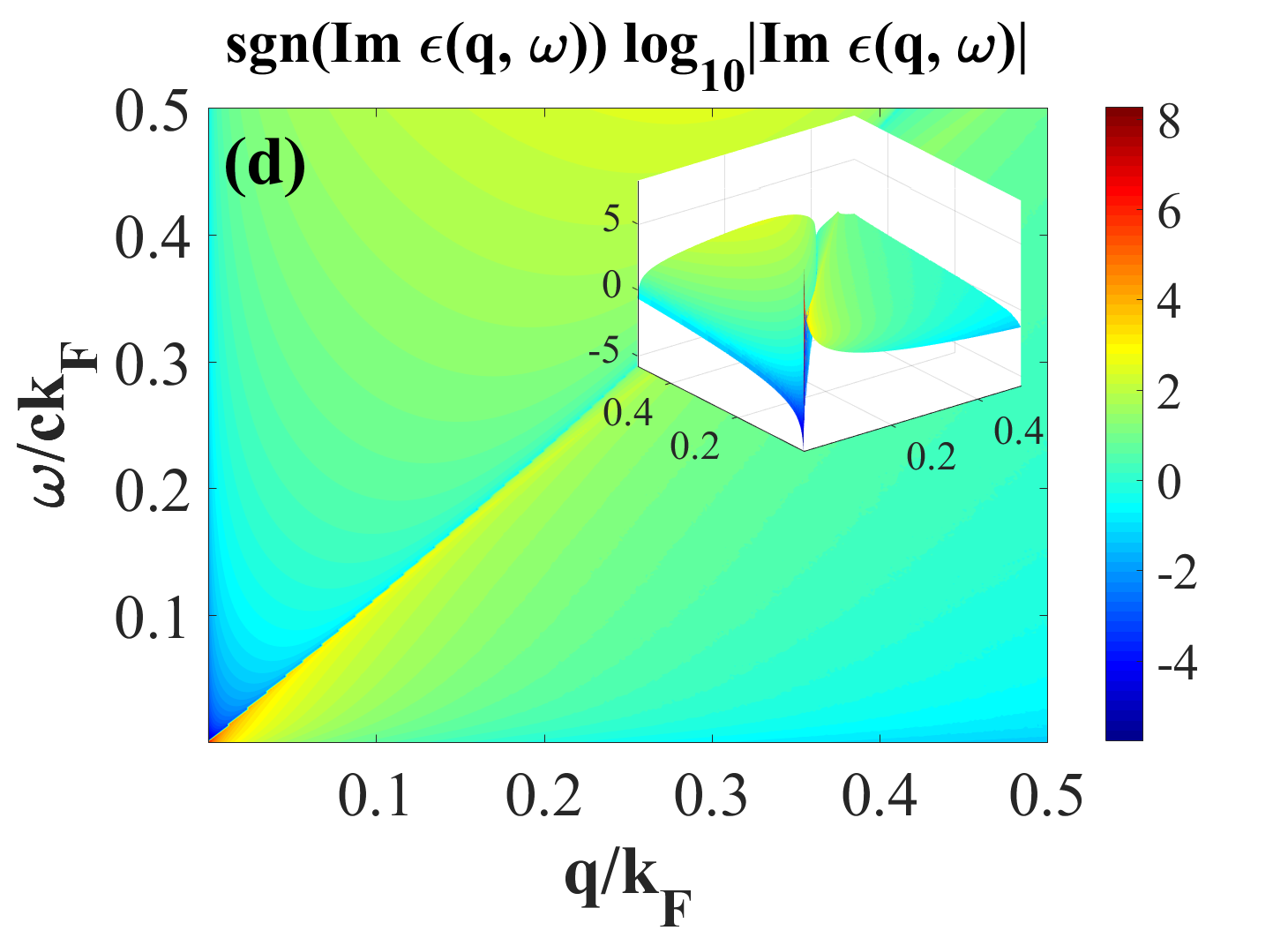}
\caption{The real (a, c) and imaginary (b, d) part of the dielectric function $\epsilon(q,\omega)$ in $q$-$\omega$ plane for a 3D gapped momentum material with the parameter $\Gamma/ck_F=4$ (a,b) and $\Gamma/ck_F=1$ (c,d) and for $T=0$. Insets show three-dimensional surface plot of $\epsilon(q,\omega)$ to illustrate additionally the behavior of the dielectric function, which is not clear visible in contour plots in the region of small $q$ and $\omega$.} 
\label{epsilon3D_T=0-001}
\end{figure}
Growth of the system dimensionality leads to enhance of the amplitude of the dielectric function oscillations and simultaneously to increase of their damping.  In the domain of small $\omega$ and $q$, the value of the real and imaginary part of $\epsilon(q,\omega)$ turns out to be significantly higher than corresponding amplitudes in the 1D and 2D cases.
For nonzero temperature ($T/ck_F=0.1$) our numerical calculations show qualitatively the same behaviour with suppression of the oscillations amplitude, as we already observed in 1D and 2D cases. 
\begin{figure}
\includegraphics[width=0.49\columnwidth]{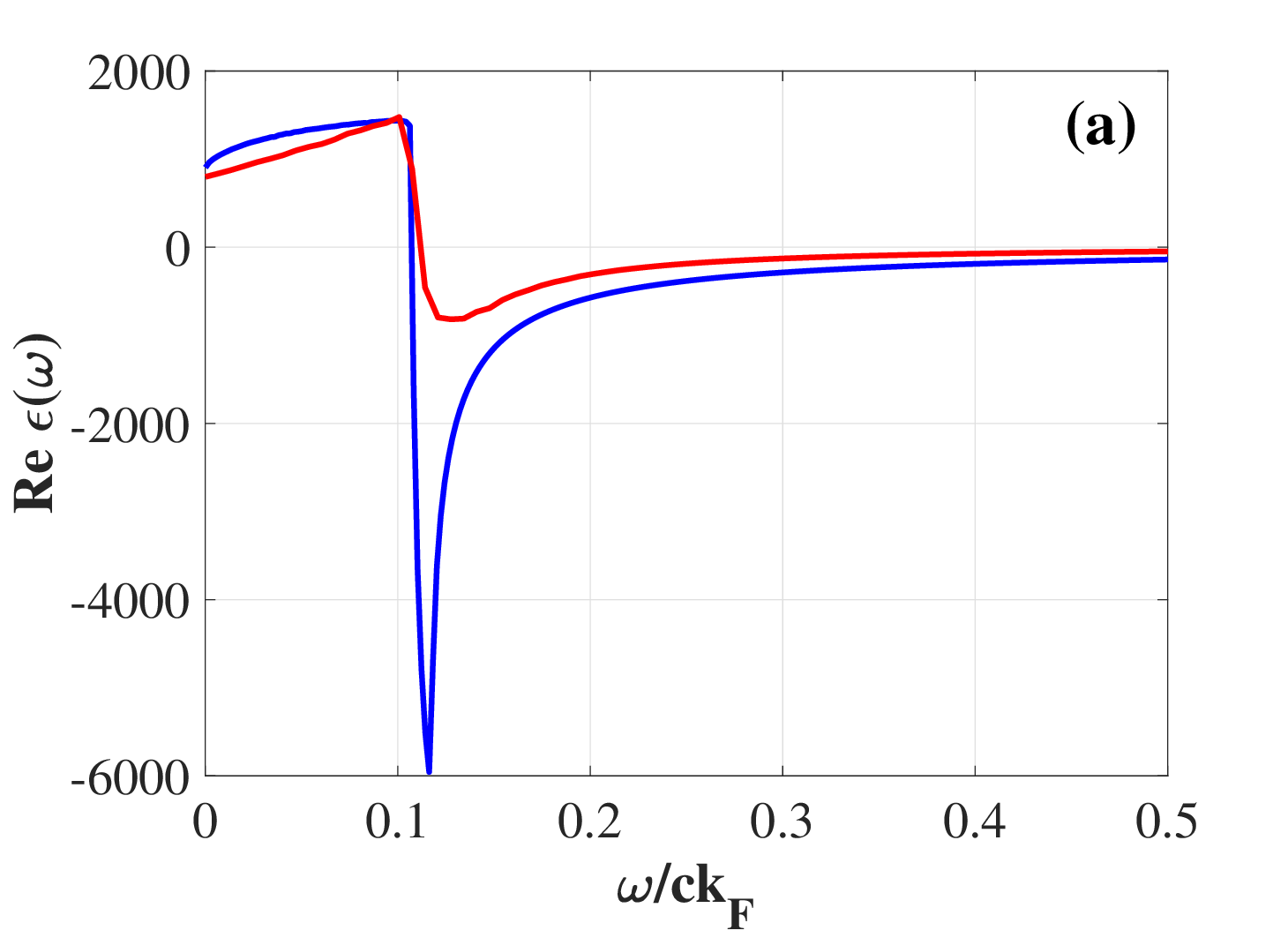}
\includegraphics[width=0.49\columnwidth]{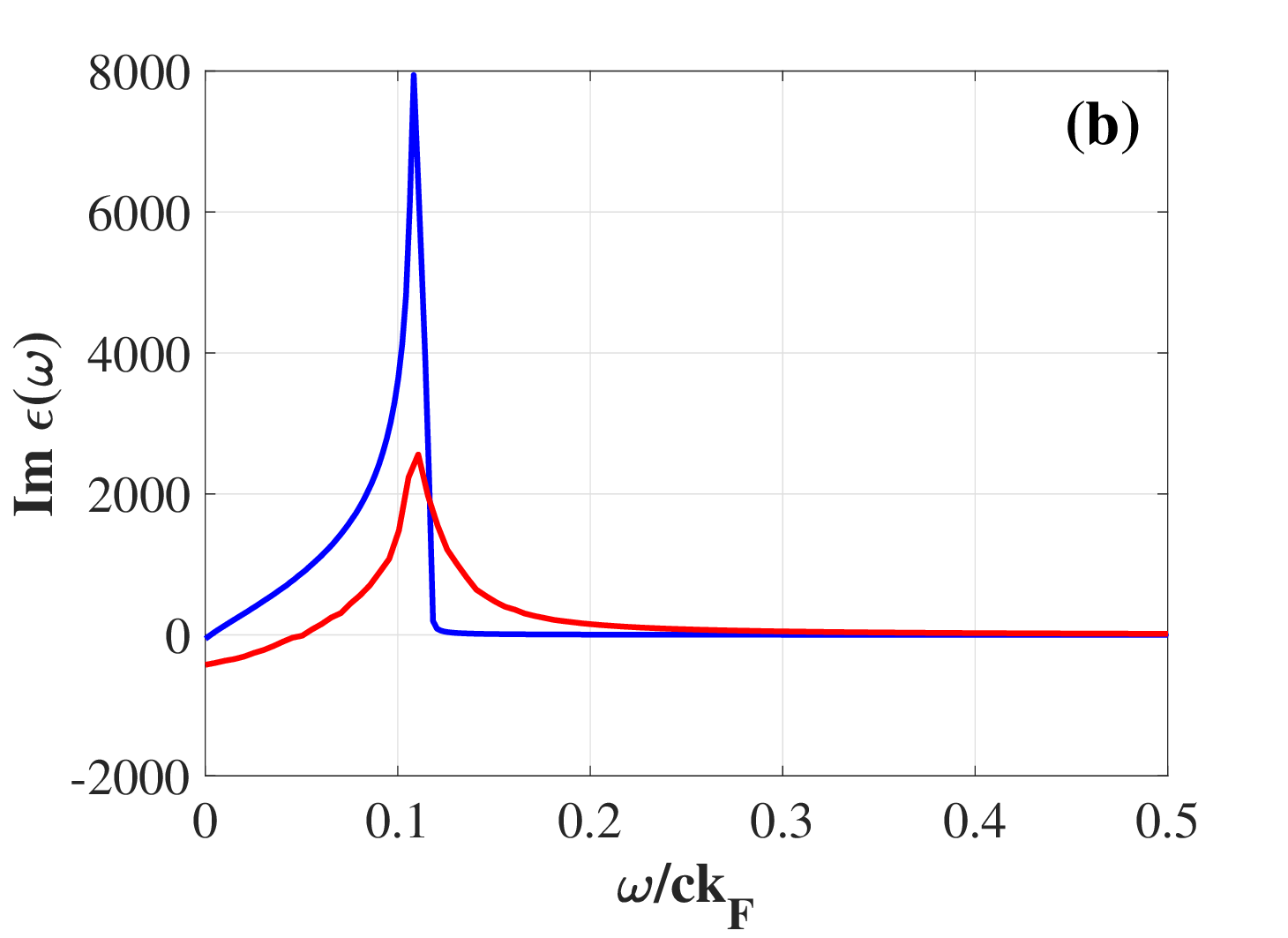}
\includegraphics[width=0.49\columnwidth]{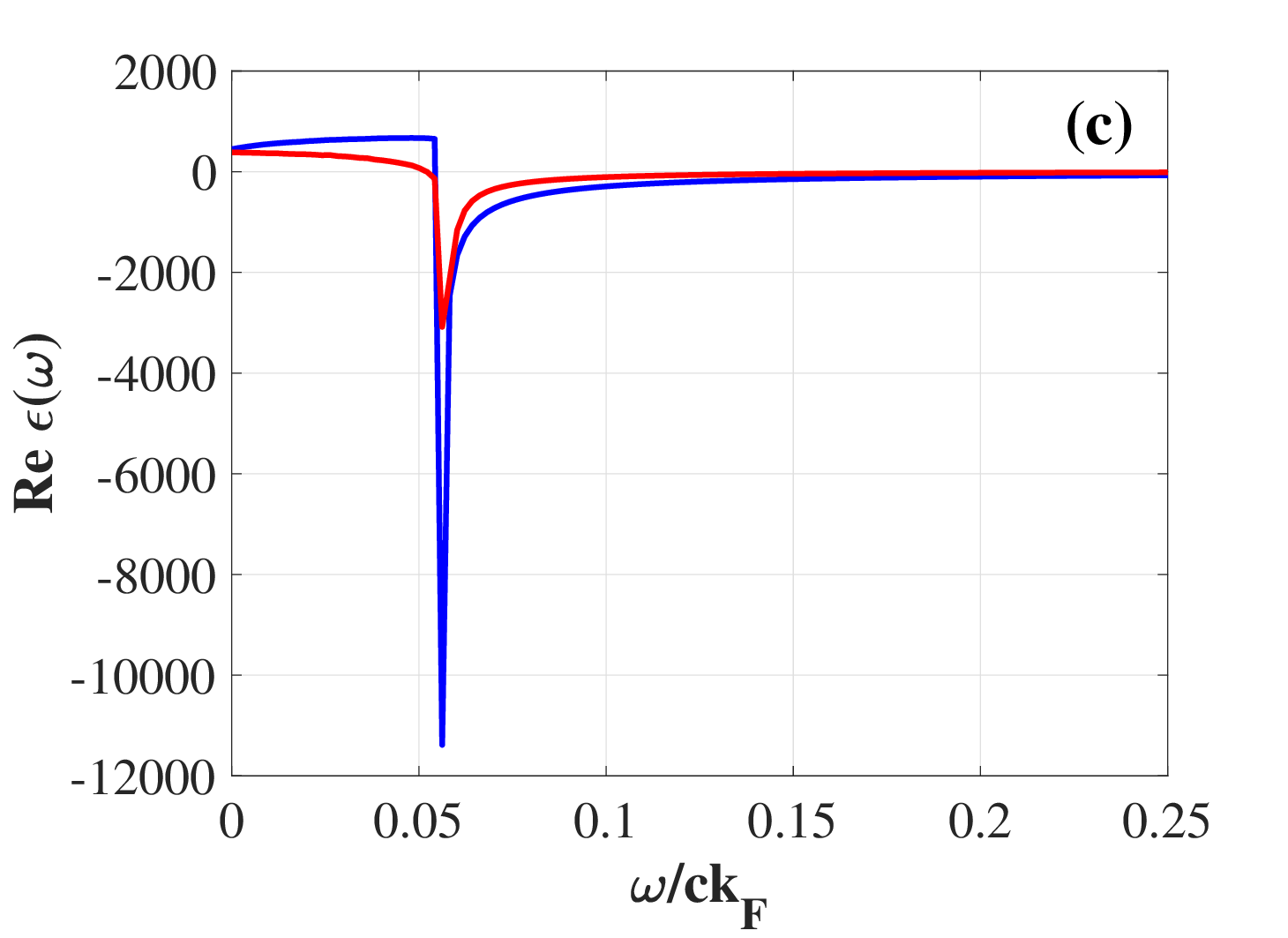}
\includegraphics[width=0.49\columnwidth]{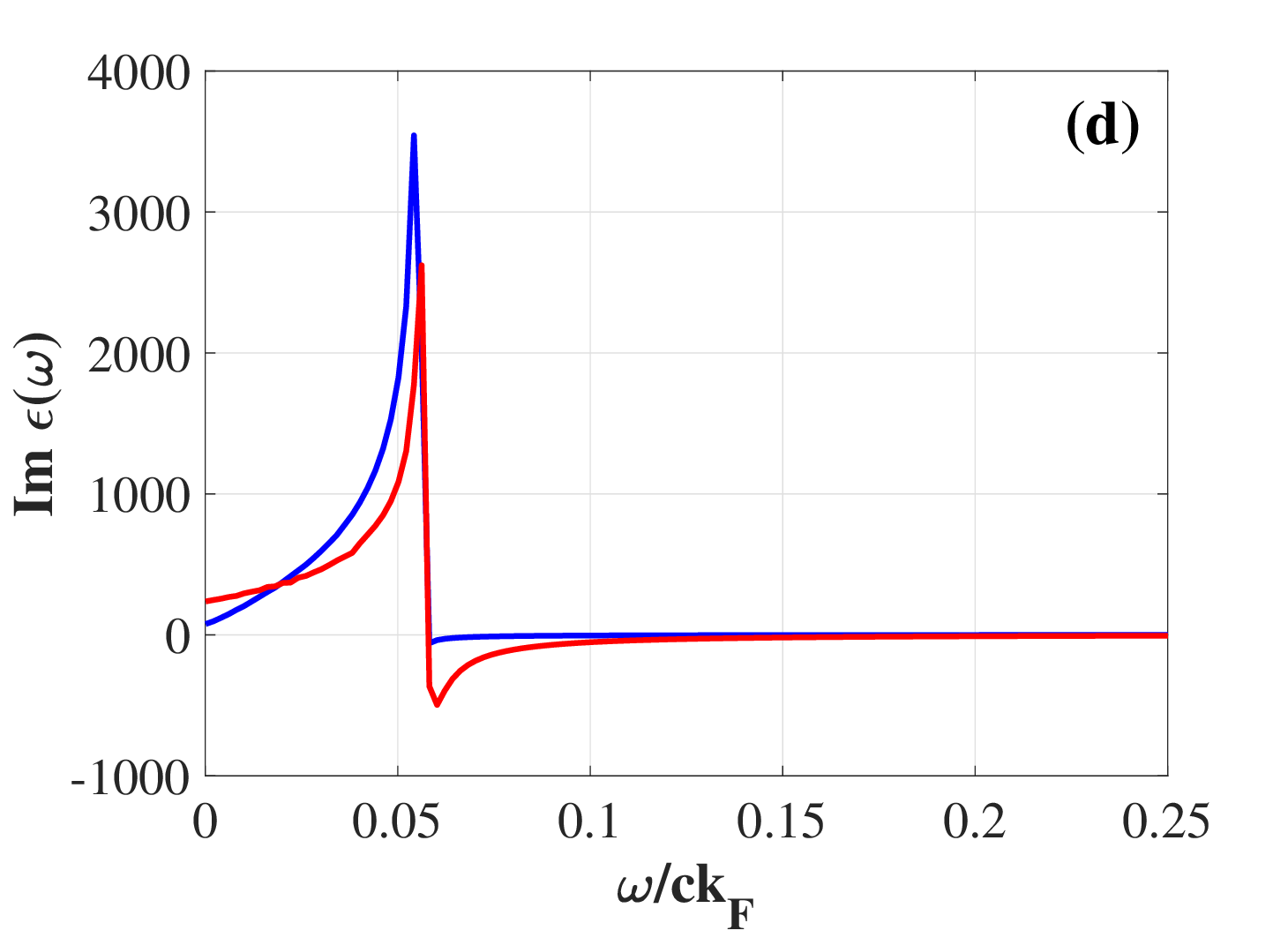}
\caption{The long-wavelength limit behavior of the the real (a, c) and imaginary (b, d) part of the dielectric finction $\epsilon(q,\omega)$ with $q=0.05k_F$ as a function of $\omega$ for a 3D gapped momentum material with the parameter $\Gamma/ck_F=4$ (a, b) and $\Gamma/ck_F=1$ (c, d) and for $T/ck_F=0$ (blue line) and $T/ck_F=0.1$ (red line).} 
\label{epsilon3D_long_wave}
\end{figure}
Figure \ref{epsilon3D_long_wave} allows to trace all mentioned features of the real and the imaginary parts of $\epsilon(q,\omega)$ for the 3D gapped momentum system that we described before. 

\subsubsection{Plasmon modes.}
The determination of plasmonic modes in the 3D case is no different from a similar procedure performed above for the low dimensional cases. Numerical solution of Eq. (\ref{plasmon_general}) at $T=0$ gives $\omega^{3D}_{pl}(q)$ shown in Figure \ref{plasmon3D_T=0-001} (a), (b) for $\Gamma/ck_F=4$ and $\Gamma/ck_F=1$, respectively.
\begin{figure}
\includegraphics[width=0.49\columnwidth]{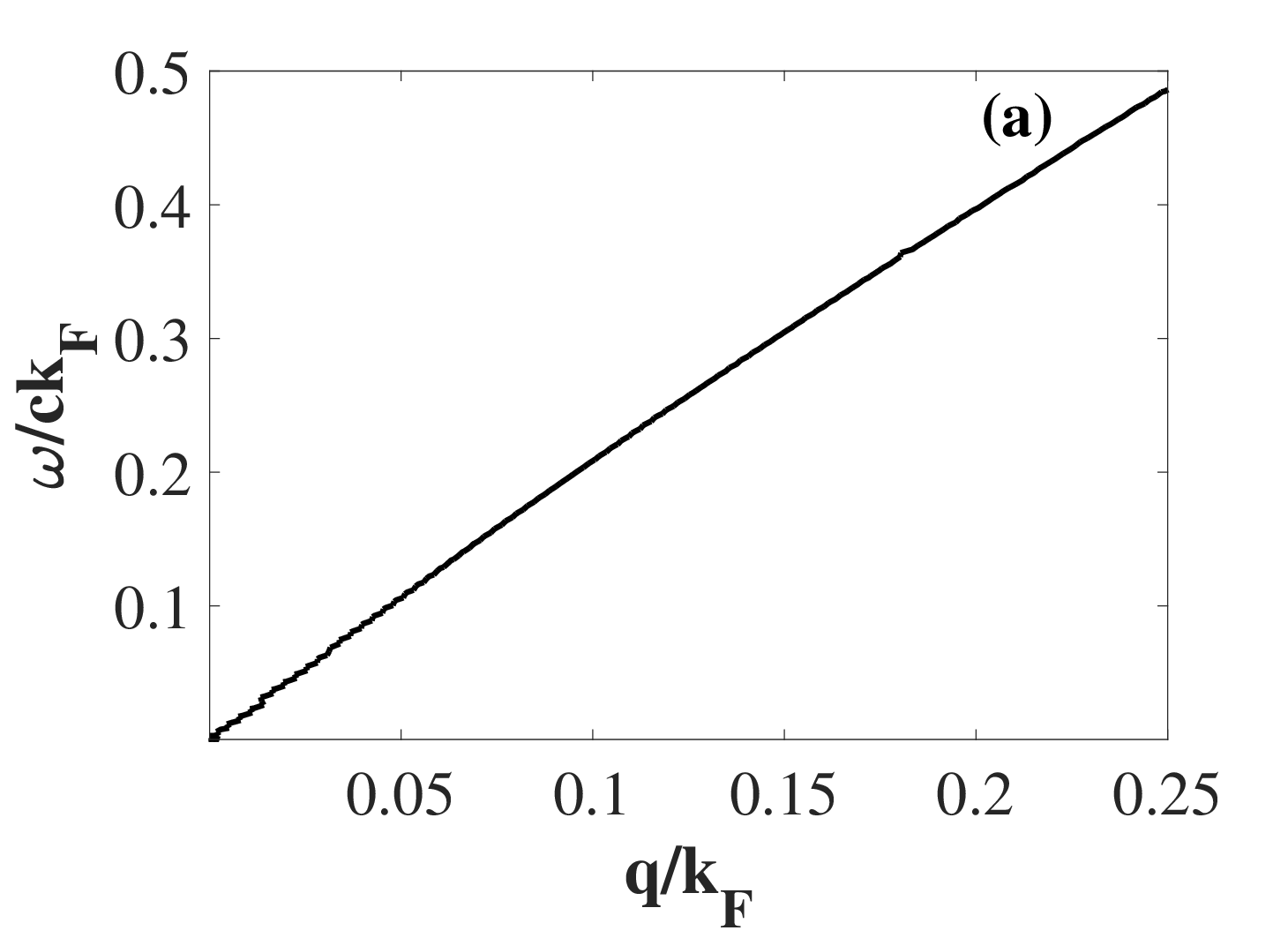}
\includegraphics[width=0.49\columnwidth]{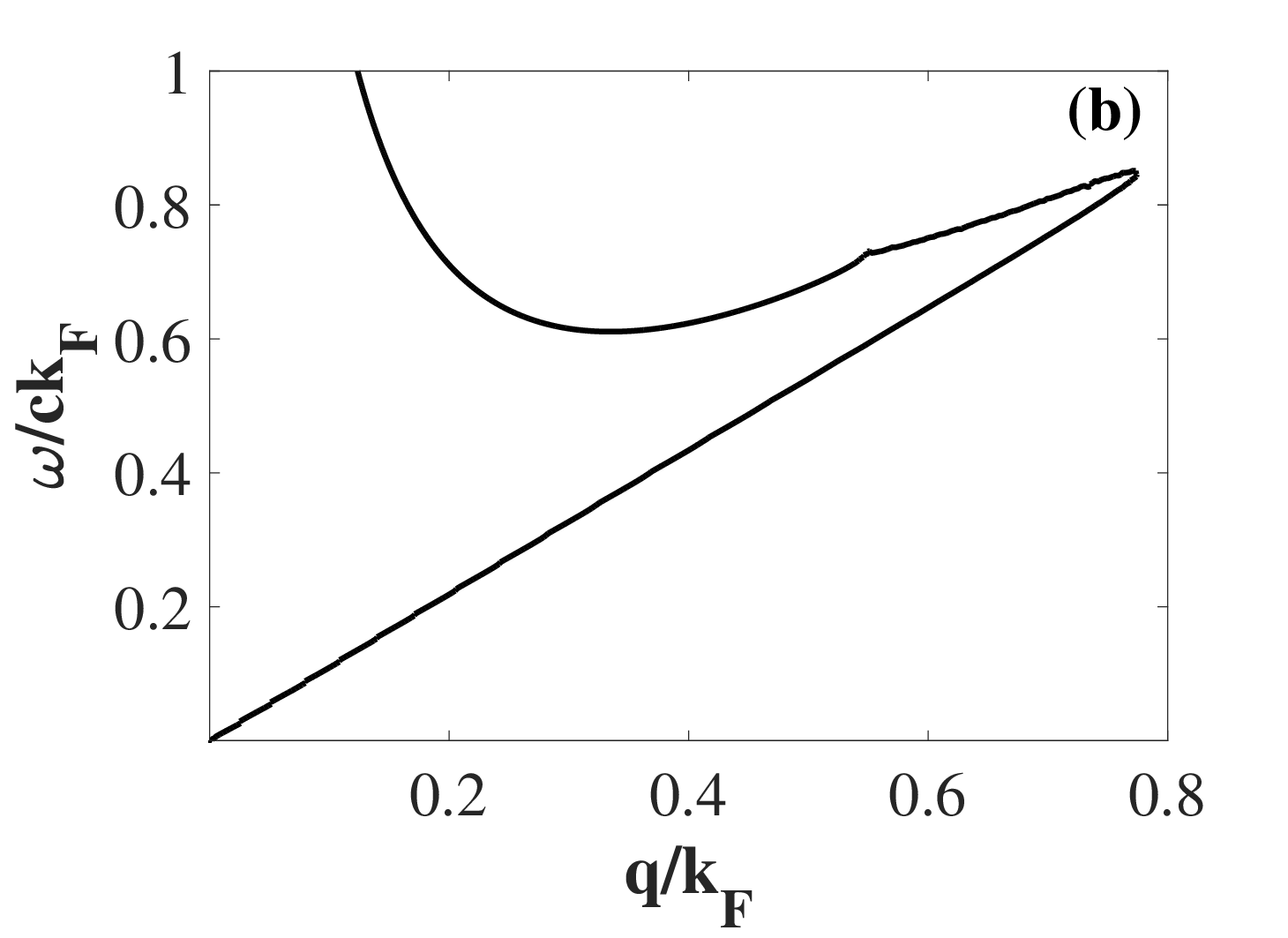}
\caption{Plasmon dispersion relations for a 3D gapped momentum material with $\Gamma/ck_F=4$ (a) and $\Gamma/ck_F=1$ (b) at low temperature $T/ck_F=0$.} 
\label{plasmon3D_T=0-001}
\end{figure}
As can be seen from Figure \ref{plasmon3D_T=0-001}, three-dimensionality imposes its own characteristic features on the function $\omega_{pl}(q)$. First of all there is no splitting and the broadening of the plasmon modes as it was observed in 1D and 2D gapped momentum system. Figure \ref{plasmon3D_T=0-001}a clearly demonstrates the linear dependence of the plasmon dispersion relation for $\Gamma/ck_F=4$. It is interesting that for another value $\Gamma/ck_F=1$, the numerical solution of Eq. (\ref{plasmon_general}) yields a unusual multi-valued function of $\omega^{3D}_{pl}(q)$ (see Figure \ref{plasmon3D_T=0-001}b). In other words, one can observe a linear lower part of the plasmon dispersion relation at low momenta with the subsequent drastic reversal in the region of large values of $q$ and its transformation into a nonlinear dependence. 

In turn, the analytical approach based on the solution of Eq. (\ref{plasmon_approx_eq}) after straightforward calculations by means of the the Sokhotski–Plemelj theorem gives the following expression for the plasmon dispersion
\begin{widetext}
\begin{equation}
\label{plasmon_approx_3D}
\omega^{3D}_{pl}(q)  = \sqrt {\frac{{2{e^2}\mu }}{{3\pi c}}\left( {\frac{{{\mkern 1mu} {\kern 1pt} \left( {2{\mkern 1mu} {\kern 1pt} {\mu ^2} + {\Gamma ^2}} \right)\sqrt {2{\mkern 1mu} {\kern 1pt} {\mu ^2} + 2{\mkern 1mu} {\kern 1pt} \sqrt {{\mu ^2}\left( {{\mu ^2} + {\Gamma ^2}} \right)} }  - \mu {\mkern 1mu} {\kern 1pt} \Gamma \sqrt {2{\mkern 1mu} {\kern 1pt} \sqrt {{\mu ^2}\left( {{\mu ^2} + {\Gamma ^2}} \right)}  - 2{\mkern 1mu} {\kern 1pt} {\mu ^2}} }}{{4{\mkern 1mu} {\kern 1pt} {\mu ^2} + {\Gamma ^2}}}} \right)},
\end{equation}
\end{widetext}
which is independent of $q$. On the qualitative level this results is in agreement with the case of 3D systems with parabolic dispersion as well as massless and gapped Dirac fermions \cite{Thakur}. However, as can be seen from the numerical solution of Eq. (\ref{plasmon_general}) (see Fig. \ref{plasmon3D_T=0-001}), this statement no longer holds true at finite $q$, where a $q$ dependence develops in the plasmon dispersion. 

\section{Absorption coefficient}

\begin{figure}
\includegraphics[width=0.49\columnwidth]{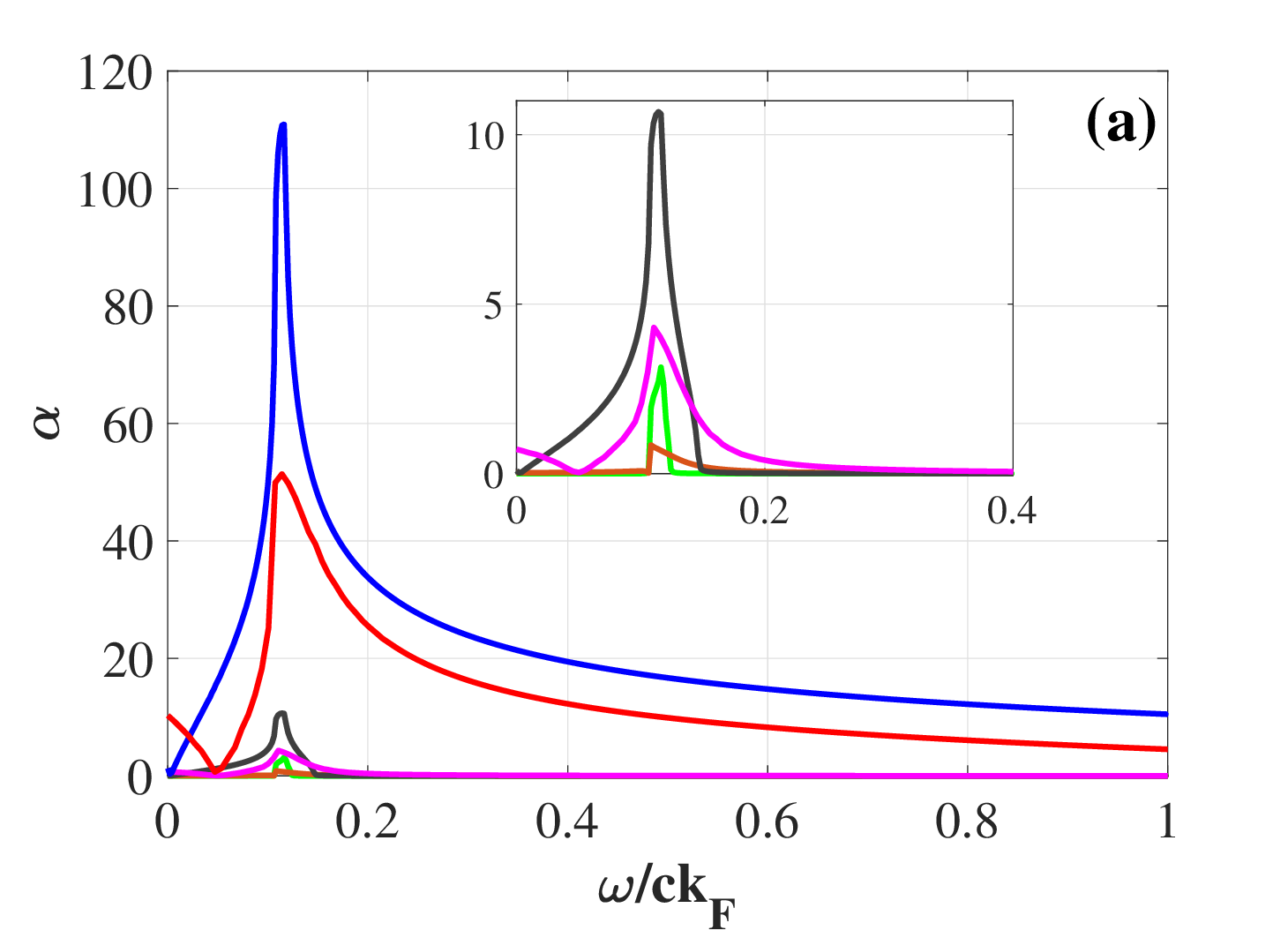}
\includegraphics[width=0.49\columnwidth]{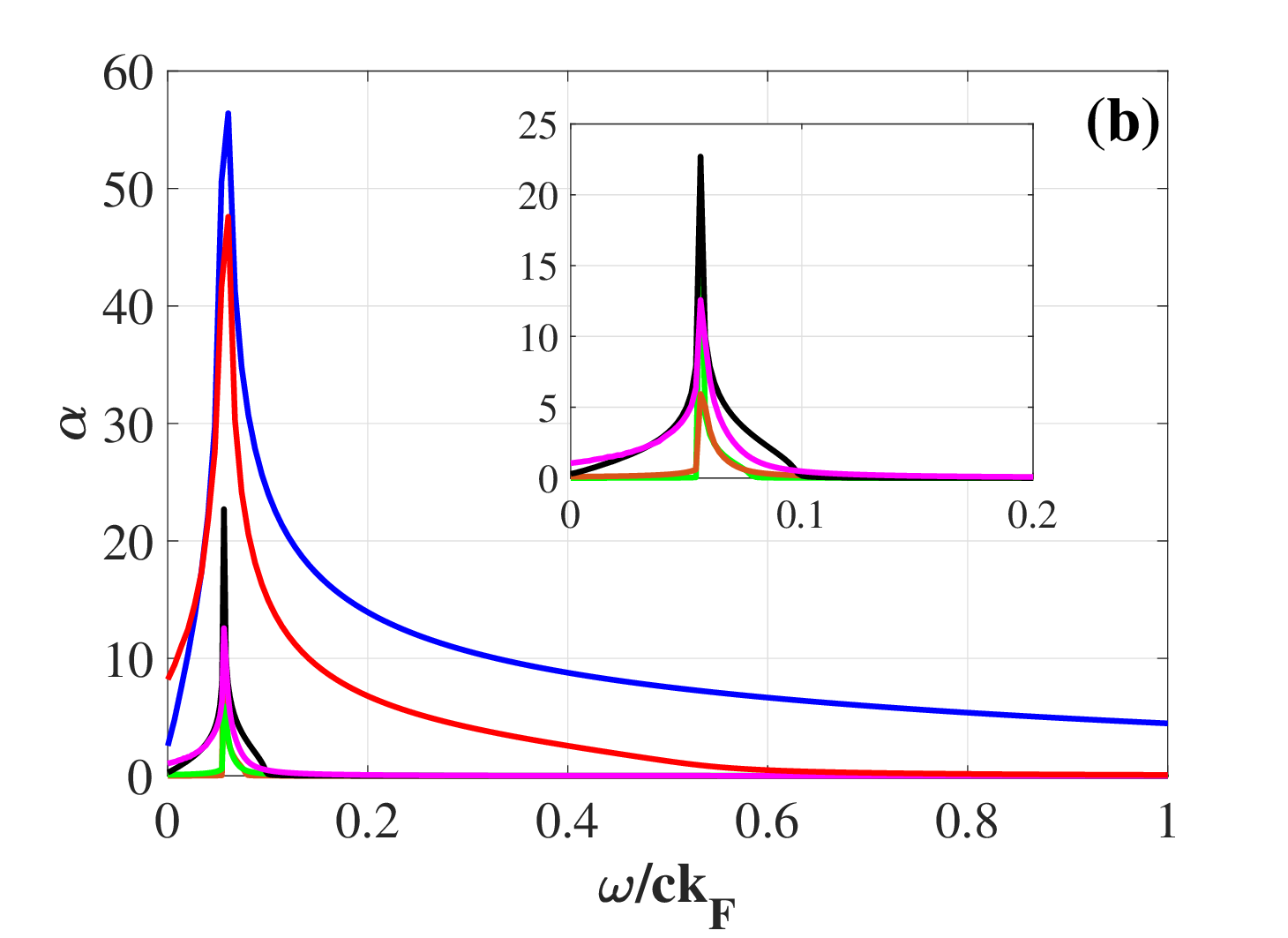}
\caption{The absorption coefficient of the 1D (green, brown lines), 2D (black, magenta lines) and 3D (blue, red lines) gapped momentum material with $\Gamma/ck_F=4$ (a) and $\Gamma/ck_F=1$ (b) at $T=0$ (blue, black, green) and $T/ck_F=0.1$ (brown, magenta, red). Insets show the zoom of the absorption coefficient dependencies for 1D and 2D cases in the vicinity of their peaks.} 
\label{absorption_plot}
\end{figure}
In this section we consider possible practical applications of the gapped momentum systems for photovoltaics. Namely, by means of the dielectric function obtained above we will study the behavior of the absorption coefficient $\alpha$ in its dependence on frequency for the systems of different dimensionalities. 

The absorption coefficient $\alpha$ can be calculated as \cite{Fox}
\begin{equation}
\label{absorption}
\alpha (q,\omega ) = \frac{{2E}}{c}\sqrt {\frac{{\sqrt {{{{\mathop{\rm Re}\nolimits} }^2}\varepsilon \left( {q,\omega } \right) + {{{\mathop{\rm Im}\nolimits} }^2}\varepsilon \left( {q,\omega } \right)}  - {\mathop{\rm Re}\nolimits} \, \varepsilon \left( {q,\omega } \right)}}{2}},
\end{equation}
where the real and imaginary parts of the dielectric function in the long-wavelength limit should be substituted to Eq. (\ref{absorption}).

The results of numerical calculations for $\alpha$ are presented in Figure \ref{absorption_plot}. It is easy to see that the higher dimensionality causes a significant increase in the absorption coefficient regardless of the temperature (blue and red lines correspond to the 3D cases for zero and nonzero temperatures, respectively). In addition, as the complex energy gap ($\Gamma$) increases, the value in the maximum of the absorption coefficient increases and it shifts to the region of lower frequencies. 

Let us make some numerical estimations for the absorption coefficient. Since the dispersion relation for the gapped momentum systems is similar to that one for gapped Dirac materials with a complex value of the energy gap, the characteristics of graphene can be utilized to estimate $\alpha$ and the frequency range, where its maximum is attained. Thus, taking into account data from Figure \ref{absorption_plot}, setting the value of the Fermi wavelength equal to 0.74 nm \cite{graphene_kf} and the Fermi velocity of the order $~{10^6}\,{\rm{m/s}}$ \cite{graphene_vf} we obtain that the absorption coefficient can reach $\alpha_{max} \approx 3.56 \cdot {10^6}\, {\rm{ c}}{{\rm{m}}^{-1}}$ for $\omega  \approx 930\,{\rm{ THz}}$, which corresponds to the soft ultraviolet range. Consequently, according to Figure \ref{absorption_plot}b the absorption coefficient should decrease by approximately a factor of 4 in the visible range of light. This value is one order of magnitude higher than the absorption coefficient of perovskites \cite{Valastro}, materials that are considered now as very promising for photovoltaic purposes. 

\section{Conclusions}

We have calculated the dielectric function of the gapped momentum material in 1D, 2D and 3D dimensions and for the case of zero and nonzero temperatures within the RPA approach. We have revealed rapidly damped oscillations of the dielectric function occurring in the domain of small frequencies $\omega$ and wave vectors $q$. The degree of the damping of these oscillations enhances with increasing of the dimensionality of the gapped momentum system, while at the same time the amplitude of both real and imaginary parts of the dielectric function increases significantly. 

What concerns the plasmon dispersion in gapped momentum systems, calculated both numerically and analytically for the long-wavelength limit, we have found the splitting of the modes and the broadening of the plasmon dispersion line in 1D and 2D (similar to the same effect observed in the 2DEG with the Rashba spin-orbit coupling and graphene). Contrary,  and such splitting is absent in 3D case, where for certain values of the parameter $\Gamma$ the plasmon dispersion transforms from the linear single-valued to the multi-valued function. Analytical calculations have shown the coincidence at the qualitative level of the plasmon dispersion relations (see Eqs. (\ref{plasmon_approx_1D}), (\ref{plasmon_approx_2D}) and (\ref{plasmon_approx_3D})) with the same characteristics for gapped Dirac materials in the long wavelength limit. 

As the practical application of our study of the dielectric properties the calculations of the absorption coefficient of gapped momentum media have been performed for 1D, 2D and 3D cases.  We found that the most significant amplification of the absorption coefficient occurs for the 3D system. Moreover, as the frequency increases, the absorption coefficient drops from its peak value less rapidly compared to the same behavior for systems of lower dimensionality. Also, based on physical characteristics of graphene as the closest material from the point of view of the dispersion law structure, we have estimated the numerical value of the absorption coefficient for gapped momentum media.

Generally speaking, one could extend our approach to solve the inverse problem: using a nonlinear integro-differential Eq. (\ref{Polarization_simple}) and assuming a given behavior of the dielectric function, the corresponding dispersion relation of a medium can be determined. In other words, this approach may pave the way for modeling of the materials with designed dielectric properties. For example, assuming that  $\epsilon(q \to 0,\omega)$ is a slowly varying function \cite{Bingham} of the frequency, based on  Eq. (\ref{Polarization_simple}), one could calculate what a dispersion relation yields this behavior. We leave this inverse problem for future studies.

We expect that this work will stimulate further theoretical and practical interest in the dielectric and other macroscopic properties of gapped momentum materials as well as of non-Hermitian systems with the complex value of the energy gap.

\begin{acknowledgments}
We acknowledge the support Mission Innovation of the MITE under the IEMAP project. A.A.V. acknowledge Quantum Matter Bordeaux (QMBx, the University of Bordeaux) for financial support and hospitality, where the part of this work was done.
\end{acknowledgments}

\begin{widetext}
\appendix
\section{Exact analytical expression for the polarization function in the 1D case}
\label{sec:A}

Here we provide the exact analytical result of the integration of the polarization function for the 1D gapped momentum system at $T=0$
\begin{equation}
\label{polarization1D_exact}
\Pi \left( {q,\omega } \right) = P\left( {{x_2},q,\omega } \right) - P\left( {{x_1},q,\omega } \right),
\end{equation}
where the function has the form
\begin{equation}
\label{P_function}
\begin{array}{l}
P\left( {x,q,\omega } \right) =  - \frac{1}{2}{\mkern 1mu} \alpha \left( x \right)\left( {{\omega ^2} - {c^2}{q^2}} \right)\sqrt {\frac{{\left( {{\Gamma ^2} + {x^2} - {c^2}{q^2}} \right)\left( {{x^2} - {c^2}{q^2}} \right)}}{{{c^2}{q^2} - {\Gamma ^2}}}} {\rm{arctanh}}\beta \left( x \right)\\
 + {\mkern 1mu} {\mkern 1mu} \frac{\omega }{{cq}}\left( {F\left( {\tilde x,m} \right) - \left( {1 - \frac{{{c^2}{q^2}}}{{{\omega ^2}}}} \right)\Pi \left( {\tilde x,\nu ,m} \right)} \right){\mathop{\rm sgn}} \left( {{\omega ^2} - {c^2}{q^2}} \right) - \frac{1}{2}{\mkern 1mu} \ln \gamma \left( x \right){\mathop{\rm sgn}} \left( {{\omega ^2} - {c^2}{q^2}} \right)\\
 + \frac{{cq\omega E\left( {\tilde x,m} \right) + \left( {\omega x - {q^2}} \right)\sqrt {\frac{{{\Gamma ^2} + {x^2} - {c^2}{q^2}}}{{{x^2} - {c^2}{q^2}}}} }}{{{\omega ^2} - {c^2}{q^2}}}{\mathop{\rm sgn}} \left( {{\Gamma ^2} + {x^2} - {c^2}{q^2}} \right)\\
 - \frac{{{c^2}{q^2}{\Gamma ^2}}}{{{\omega ^2} - {c^2}{q^2}}}\left( {\frac{1}{2}\frac{{{\mkern 1mu} {\rm{arctanh}}\beta \left( x \right)}}{{\sqrt {\left( {{\omega ^2} - {c^2}{q^2}} \right)\left( {{\Gamma ^2} + {\omega ^2} - {c^2}{q^2}} \right)} }} + \frac{{\Pi \left( {\tilde x,\nu ,m} \right)}}{{\omega cq}}} \right){\mathop{\rm sgn}} \left( {{\omega ^2} - {c^2}{q^2}} \right),
\end{array}
\end{equation}
where 
\begin{equation}
\label{x1_x2}
{x_1} = \frac{{i\Gamma }}{2} - \sqrt {{c^2}{q^2} - \frac{{{\Gamma ^2}}}{4}},
{x_2} = \sqrt {{c^2}k_F^2 - \frac{{{\Gamma ^2}}}{4}}  - \sqrt {{c^2}{{\left( {{k_F} + q} \right)}^2} - \frac{{{\Gamma ^2}}}{4}},
\end{equation}
and where we introduce new variables and additional functions
\begin{equation}
\label{alpha}
\alpha \left( x \right) = \sqrt {\frac{{{c^2}{q^2} - {\Gamma ^2}}}{{\left( {{\Gamma ^2} + {x^2} - {c^2}{q^2}} \right)\left( {{\Gamma ^2} + {\omega ^2} - {c^2}{q^2}} \right)\left( {{x^2} - {c^2}{q^2}} \right)\left( {{\omega ^2} - {c^2}{q^2}} \right)}}},
\end{equation}
\begin{equation}
\label{beta}
\beta \left( x \right) = \frac{1}{2}\frac{{{\Gamma ^2}\left( {{\omega ^2} - 2{c^2}{q^2} + {x^2}} \right){\mkern 1mu}  + 2\left( {{x^2} - {c^2}{q^2}} \right)\left( {{\mkern 1mu} {\omega ^2} - {c^2}{q^2}{\mkern 1mu} } \right)}}{{\sqrt {\left( {{\Gamma ^2} + {x^2} - {c^2}{q^2}} \right)\left( {{\Gamma ^2} + {\omega ^2} - {c^2}{q^2}} \right)\left( {{x^2} - {c^2}{q^2}} \right)\left( {{\mkern 1mu} {\omega ^2} - {c^2}{q^2}{\mkern 1mu} } \right)} }},
\end{equation}
\begin{equation}
\label{gamma}
\gamma \left( x \right) = {\Gamma ^2} + 2{\mkern 1mu} \left( {{x^2} - {c^2}{q^2}} \right) + 2{\mkern 1mu} \sqrt {\left( {{\Gamma ^2} + {x^2} - {c^2}{q^2}} \right)\left( {{x^2} - {c^2}{q^2}} \right)},
\end{equation}
\begin{equation}
\label{new_x}
\tilde x = \frac{x}{{\sqrt {{c^2}{q^2} - {\Gamma ^2}} }},
\end{equation}
\begin{equation}
\label{elliptic_k}
m = \frac{{\sqrt {{c^2}{q^2} - {\Gamma ^2}} }}{{cq}},
\end{equation}
\begin{equation}
\label{elliptic_nu}
\nu  = \frac{{{c^2}{q^2} - {\Gamma ^2}}}{{{\omega ^2}}}.
\end{equation}

\section{Exact analytical expression for the polarization function in the 2D case}
\label{sec:B}
For the 2D case we can perform the integration over polar angle $\varphi$ only
\begin{equation}
\label{polarization2D_exact}
\begin{array}{l}
\Pi \left( {q,\omega } \right) = \int\limits_0^{{k_F}} {kdk} \int\limits_0^{2\pi } {\frac{{d\varphi }}{{\omega  + \sqrt {{c^2}{k^2} - \frac{{{\Gamma ^2}}}{4}}  - \sqrt {{c^2}{k^2} + {c^2}{q^2} - \frac{{{\Gamma ^2}}}{4} + 2{c^2}kq\cos \varphi } }}} \\
 =  - \int\limits_0^{{k_F}} {\frac{{4kdk{\mkern 1mu} }}{{\left( {{{\left( {\omega  + \sqrt {{c^2}{k^2} - \frac{{{\Gamma ^2}}}{4}} } \right)}^2} - \left( {{c^2}{{\left( {k - q} \right)}^2} - \frac{{{\Gamma ^2}}}{4}} \right)} \right)\sqrt {{c^2}{{\left( {k - q} \right)}^2} - \frac{{{\Gamma ^2}}}{4}} }}} \\
\left[ {\left( {{{\left( {\omega  + \sqrt {{c^2}{k^2} - \frac{{{\Gamma ^2}}}{4}} } \right)}^2} - \left( {{c^2}{{\left( {k - q} \right)}^2} - \frac{{{\Gamma ^2}}}{4}} \right)} \right){\rm{K}}\left( {2c\sqrt {\frac{{kq}}{{\frac{{{\Gamma ^2}}}{4} - {c^2}{{\left( {k - q} \right)}^2}}}} } \right)} \right.\\
 - \left. {{{\left( {\omega  + \sqrt {{c^2}{k^2} - \frac{{{\Gamma ^2}}}{4}} } \right)}^2}\Pi \left( {\frac{{4{c^2}kq}}{{{{\left( {\omega  + \sqrt {{c^2}{k^2} - \frac{{{\Gamma ^2}}}{4}} } \right)}^2} - \left( {{c^2}{{\left( {k - q} \right)}^2} - \frac{{{\Gamma ^2}}}{4}} \right)}},2c\sqrt {{\mkern 1mu} \frac{{kq}}{{\frac{{{\Gamma ^2}}}{4} - {c^2}{{\left( {k - q} \right)}^2}}}} } \right)} \right].
\end{array}
\end{equation}

Further analytical integration is not possible.

\section{Exact analytical expression for the polarization function in the 3D case}
\label{sec:C}
For the 3D case we can perform the integration over polar and azimuthal angles $\varphi$ and $\theta$, respectively
\begin{equation}
\label{polarization3D_exact}
\begin{array}{l}
\Pi \left( {q,\omega } \right) =  - 2\pi \int\limits_0^{{k_F}} {\frac{{k\left( {\omega  + \sqrt {{c^2}{k^2} - \frac{{{\Gamma ^2}}}{4}} } \right)}}{{{c^2}q}}}  \times \\
\left[ {\ln \left( {\frac{{ - \left( {\omega  + \sqrt {{c^2}{k^2} - \frac{{{\Gamma ^2}}}{4}} } \right) + \sqrt {{c^2}{{\left( {k + q} \right)}^2} - \frac{{{\Gamma ^2}}}{4}} }}{{ - \left( {\omega  + \sqrt {{c^2}{k^2} - \frac{{{\Gamma ^2}}}{4}} } \right) + \sqrt {{c^2}{{\left( {k - q} \right)}^2} - \frac{{{\Gamma ^2}}}{4}} }}} \right) + \sqrt {{c^2}{{\left( {k + q} \right)}^2} - \frac{{{\Gamma ^2}}}{4}}  - \sqrt {{c^2}{{\left( {k - q} \right)}^2} - \frac{{{\Gamma ^2}}}{4}} } \right]dk,
\end{array}
\end{equation}
where the integral over $k$ cannot be represented in the analytic form.
\end{widetext}

\end{document}